\documentclass[12pt]{article}
\usepackage{graphicx}
\usepackage{epsfig}
\usepackage{rotating}
\usepackage{dcolumn}
\usepackage{bm}
\usepackage{cite}%
\newcommand{\comment}[1]{}
\newcommand{\nn}{\nonumber}
\newcommand{\beq}{\begin{equation}}
\newcommand{\eeq}{\end{equation}}
\newcommand{\bea}{\begin{eqnarray}}
\newcommand{\eea}{\end{eqnarray}}
\newcommand{\barr}{\begin{array}}
\newcommand{\earr}{\end{array}}
\def\dis{\displaystyle}
\newcommand{\fb}{\, {\rm fb}}
\newcommand{\gsim}
{{\;\raise0.3ex\hbox{$>$\kern-0.75em\raise-1.1ex\hbox{$\sim$}}\;}}

\newcommand{\lsim}
{{\;\raise0.3ex\hbox{$<$\kern-0.75em\raise-1.1ex\hbox{$\sim$}}\;}}

\def\lesq{\ell_e^2}
\def\resq{r_e^2}

\def\lfsq{\ell_f^2}
\def\rfsq{r_f^2}

\def\ltsq{\ell_\tau^2}
\def\rtsq{r_\tau^2}
\begin{document}

\vspace*{-1em}
\begin{flushright}
{IISc-CHEP/02/08}
\end{flushright}
\vspace*{1cm}
\begin{center}
\textbf{\Large Role of polarization in probing anomalous gauge interactions
of the Higgs boson}
\end{center}
\begin{center}
{\mbox{Sudhansu~S.~Biswal$^1$,
Debajyoti Choudhury$^2$,
Rohini~M.~Godbole$^1$ 
and Mamta$^3$}} \\ \vspace*{0.2cm}
{\small \it $^1$Centre for High Energy Physics, Indian Institute of Science,
Bangalore 560 012, India}\\
{\small \it $^2$Department of Physics and Astrophysics,
University of Delhi, Delhi 110 007, India}\\
{\small \it $^3$Department of Physics and Electronics, S.G.T.B. Khalsa College,}\\
{\small \it University of Delhi, Delhi 110 007, India}
\end{center}

\begin{abstract}
We explore the use of polarized $e^+/e^-$ beams and/or the information
on final state decay lepton polarizations in probing the interaction
of the Higgs boson with a pair of vector bosons.  A model independent
analysis of the process $e^+ e^- \rightarrow f \bar f H$, where $f$ is
any light fermion, is carried out through the construction of observables
having identical properties under the discrete symmetry
transformations as different individual anomalous interactions.  
This allows us to probe an individual anomalous term independent of the
others.  We find that initial state beam polarization can significantly 
improve the sensitivity to $CP$-odd couplings of the $Z$ boson 
with the Higgs boson ($ZZH$). 
Moreover, an ability to 
isolate events with a particular $\tau$ helicity, with even  40\%
efficiency, can improve sensitivities to certain $ZZH$ couplings 
by as much as a factor of 3. In addition, the contamination from the $ZZH$ 
vertex contributions present in the measurement of the 
trilinear Higgs-$W$ ($WWH$) couplings can be  
reduced to a great extent by employing  polarised beams. The effects of 
initial state radiation (ISR) 
and beamstrahlung, which can be relevant for higher values of the beam
energy are also included in the analysis.
\end{abstract}

\section{\label{sec:intro}Introduction}

Although the Standard Model (SM) has been highly successful in
describing all the available experimental data, the precise mechanism
of the breaking of the $\rm {SU(2)}\otimes \rm{U(1)}$ gauge symmetry
and consequent generation of masses for all the elementary particles
is still very much an open question. In the SM, the symmetry is broken
spontaneously giving rise to masses for all the elementary particles
via the Higgs mechanism thereby requiring the presence of a spin-0
CP-even particle, namely the Higgs
boson~\cite{Higgs,HHG,maggie,abdelrev}.  However, so far, there exists
no direct experimental evidence for the the same.  Not surprisingly,
therefore, the search for the Higgs boson and the study of its various
properties, comprise one of the major aims of all the current and
future colliders~\cite{Godbole:2002mt}. The Large Hadron Collider
(LHC), soon to go into operation, is expected to shed light on the
mechanism of electroweak symmetry breaking (EWSB). It is designed to
be capable of finding the SM Higgs boson over most of the
theoretically allowed range for its mass~\cite{LHC}.

Direct searches at the LEP gives a lower bound on the mass of the SM
Higgs boson: $m_H > 114.4$ GeV~\cite{higmin}. On the other hand,
electroweak precision measurements put an upper bound on its mass of
about 182 GeV at the 95\% confidence level (CL)~\cite{higmax}. Note,
though, that these mass bounds are model-dependent and various
extensions of the SM admit different allowed ranges of $m_H$.  For
example, the lower bound can be relaxed in generic two-Higgs doublet
models (2HDM)~\cite{2hdm} and more spectacularly in multi-Higgs models
with CP violation~\cite{cp-mssm,cp-multihiggs}. As a matter of fact,
even in the Minimal Supersymmetric extension of the Standard Model
(MSSM)~\cite{MSSMbook}, once additional CP-violation is admitted in the scalar
sector~\cite{cp-mssm}, direct searches at LEP and elsewhere still
allow a Higgs boson mass as low as 10 GeV~\cite{lowerH}.  Similarly, 
the non-minimal supersymmetric standard model admits very light 
spin-0 states even without invoking additional sources of $CP$ 
violation~\cite{cpnsh}. In certain
extensions, the upper bound on the mass of the (lightest) Higgs boson
may also be substantially higher~\cite{highH}. A more detailed
discussion of the subject may be found in Refs.~\cite{abdelrev,cpnsh}.

While the SM contains only a single CP-even scalar state, in general,
various extensions of the SM mentioned above contain more than one 
Higgs boson and 
some, possibly, with different $CP$ properties.  For example, the
2HDM---of which the MSSM is a particular
case---consists of five spin zero particles: two CP-even neutrals, a
CP-odd neutral and a pair of charged scalars. If the MSSM parameters
admit CP violation, the neutral particles may no longer be CP
eigenstates. The aforementioned dilution of the experimental lower
limits is, generically, the result of a reduced coupling of the
lightest spin-0 state with the $Z$ due to the mixing of the SM higgs
with the other (pseudo)scalars in the model.

Thus, even after the LHC sees a signal for a Higgs boson, a study of
its properties (including CP) and precise measurements of its
interactions would be necessary to establish the nature of electroweak
symmetry breaking.  Such a detailed study of this sector may also
provide the footprints of new physics beyond the SM.  This, though,
would be possible only at the International $e^+e^-$ Linear Collider
(ILC)~\cite{ILC,RDR} and combined information from the ILC and
LHC~\cite{Weiglein:2004hn,Godbole:2004xe}, will be required to
establish it as {\em the} SM Higgs boson. A key step in this direction
is the determination of the tensor structure of the coupling of the
spin-$0$ state with the different SM particles.  A model independent
analysis would, then, incorporate the most general form for this
tensor structure as allowed by symmetry principles, the anomalous
parts having been assumed to have come from effects of high scale
physics.
 The couplings of the Higgs boson with 
a pair of gauge bosons $V (V = \gamma, W$ and Z) as well as that with a 
$t \bar t$ pair have been studied very thoroughly in the context.  The 
tensor structure can be inferred from kinematical distributions 
and polarisation measurements for various final state particles. 
In this study, we concentrate on 
the trilinear Higgs-$Z$ ($ZZH$) and Higgs-$W$ ($WWH$) coupling, 
in particular 
focusing on the utility of beam polarisation, measurement of the final state 
particle polarisation as well as the use of higher beam energies, for the 
process $e^+ e^- \rightarrow f \bar f H$.

At an $e^+e^-$ collider, the $Z$ boson produced in the process $e^+
e^- \rightarrow Z H$ is, at high energies, longitudinally polarised
when produced in association with a CP-even Higgs boson and
transversely polarised in case of a CP-odd Higgs boson. The angular
and energy distribution of the $Z$ boson can, thus, provide a wealth of
information about the $ZZH$ coupling~\cite{Barger:1993wt, CP-full}.
The shapes of the threshold excitation curve in the processes $e^+e^-
\rightarrow Z H$~\cite{Miller:2001bi,Dova:2003py} and $e^+ e^-
\rightarrow t \bar t H$~\cite{bhupal} constitute model independent probes
of the tensor structure of the $ZZH$ and the $t \bar t H$ coupling
respectively.  Many detailed studies of how kinematical distributions
for the processes $e^+ e^- \to f \bar f H$, proceeding via vector
boson fusion and Higgs-strahlung can be used to probe the $ZZH$
vertex exist~\cite{dist,Hagiwara:1993sw}. The anomalous $ZZH$ vertex,
in the context of higher dimensional operators has been studied in
Refs.~\cite{ Hagiwara:1993sw,Rattazzi:1988ye,He:2002qi,
Hagiwara:1993ck,
Gounaris:1995mx,Hagiwara:2000tk,Chang:1993jy,Han:2000mi,Biswal:2005fh,
Biswal:2007zz}
for a Linear Collider (LC).  Ref.~\cite{Hagiwara:2000tk} is one of the
pioneering studies and contains a very extensive analysis, using the
optimal observable technique~\cite{Atwood:1991ka}, to probe $ZZH$ and
$\gamma ZH$ couplings, whereas Refs.~\cite{Chang:1993jy,
Han:2000mi,Biswal:2005fh} use asymmetries constructed using
differences in the kinematical distributions of the decay products.

The  $ZZH$ vertex could  be probed at the LHC in a similar 
fashion, again using kinematic distributions, threshold behaviour as well
as asymmetries in
the Higgs decays~\cite{Choi:2002jk,Buszello:2002uu,Bluj,Allanach:2006yt,
Zhang:2003it,Lietti:2000gg,Godbole:2007cn}. This, alongwith
 the $WWH$ vertex, can also be studied through vector boson fusion at
the LHC~\cite{Plehn:2001nj,Buszello:2006hf}.  Angular distributions of
the decay products have been used in Ref.~\cite{Niezurawski:2004ga} to
study the $VVH$ (where $V = Z/W$) vertex in the process 
$\gamma \gamma \to H \to W^+ W^- / ZZ$.

\comment{Kinematical distributions for the processes $e^+ e^- \to f
\bar f H$, proceeding via vector boson fusion and Higgsstrahlung, have
been studied both without~\cite{dist} and with beam
polarization~\cite{Romao:1986cz}. In the context of higher dimensional
operators, the anomalous $ZZH$ couplings have been studied in
Refs.~\cite{ Rattazzi:1988ye, Hagiwara:1993sw, He:2002qi,
Hagiwara:1993ck, Gounaris:1995mx} for the LC and in
Refs.~\cite{Zhang:2003it, Plehn:2001nj, Lietti:2000gg, Godbole:2007cn}
for the LHC.  Ref.~\cite{Hagiwara:2000tk} uses the optimal observable
technique~\cite{Atwood:1991ka} to probe $ZZH$ and $\gamma ZH$
couplings, whereas Refs.~\cite{Biswal:2005fh, Han:2000mi} use
asymmetries in kinematical distributions. Angular distributions of the
decay products have been used in Ref.~\cite{Niezurawski:2004ga} to
study the $VVH$ vertex in the process $\gamma \gamma \to H \to W^+ W^-
/ ZZ$. In Ref.~\cite{Biswal:2005fh}, the sensitivity to $VVH$ ($V =
Z/W$) interactions at the ILC is discussed for unpolarized
beams. Several observables (like various asymmetries) with specific
transformation properties under the discrete transformations $C,\,P$
and $\tilde T$ are constructed to probe different anomalous couplings
possible.  }

In Ref.~\cite{Biswal:2005fh}, an exhaustive set of asymmetries,
which could probe each of the $ZZH$ anomalous coupling
independent of the others,  were
constructed.  Defining kinematical observables which
are either odd or even under the different discrete symmetry
transformations, the said asymmetries are the expectation values of
the sign of these observables. In the approximation of small
contribution from anomalous parts (which amounts to retaining terms
only upto the linear order in the anomalous couplings), these
asymmetries are then proportional to the coefficient of the term in
the Lagrangian with the corresponding transformation
properties. However, many of these asymmetries turned out to be
proportional to the difference between the squared right and left
handed couplings of the fermion to the $Z$ boson (viz. $l_f^2-r_f^2$),
and consequently were rather small, on account of the electrons
(charged leptons) being involved. It follows then that the sensitivity
of these asymmetries to the anomalous couplings could be enhanced by
either using the polarized beams or through a measurement of the
polarization of the final state particles.

In the unpolarized case, the determination of the anomalous $WWH$
coupling suffers a large contamination from the contribution from the
$s$--channel diagram, arising from the $ZZH$ coupling.  We look at the
possibility of reducing this contamination by the use of polarised
beams. For completely polarized $e^+$ and $e^-$ beams, $\sigma_{LR}$
receives contributions from both the Bjorken ($s$-channel) and fusion
($t$-channel) diagrams, whereas only the $s$-channel diagram
contributes to $\sigma_{RL}$.  Thus, beam polarisation may also be
used to enhance the sensitivity to $WWH$ anomalous couplings, and this
constitutes part of our investigations.  Furthermore, we also study
the dependence of the sensitivities on the beam energy; again with an
aim to enhance the $t$ channel contribution and hence the sensitivity
to the $WWH$ couplings.  Effects of both initial state radiation
(ISR)~\cite{Kuraev:1985hb} and
beamstrahlung~\cite{Chen:1991wd,Drees:1992ws} have been included in
this study as they ought to be.

The rest of the paper is organized as follows: in Sec.~\ref{sec:vvh} we
discuss possible sources of anomalous $VVH$ couplings. Various
kinematical cuts on final state particles used to suppress the
background are discussed in Sec.~\ref{sec:cuts}. The $ZZH$ vertex is
examined in detail in Sec.~\ref{sec:zzh}, with the various observables 
being defined in Sec.~\ref{sec:obs}, the effects of beam
polarization being discussed in Sec.~\ref{sec:beamzzh} and the use of
final state $\tau$-polarization in Sec.~\ref{sec:finaltau}.  In
Sec.~\ref{sec:beamtau} we discuss the improvements possible in the
reach for the anomalous $ZZH$ coupling using final state
$\tau$ measurement with polarised initial beams.  In Sec.~\ref{sec:wwh}
we construct some observables using the polarisation of initial
beams to constrain the $WWH$ couplings. In Sec.~\ref{sec:sensitivity}
we present results on the dependence of the sensitivities to different
couplings to the beam energy, including the effects of ISR and
beamstrahlung.  Finally we summarize our results in
Sec.~\ref{sec:conclude}.

\section{\label{sec:vvh}The VVH Couplings}
Within the SM/MSSM, the only interaction term
involving the Higgs boson and a pair of gauge bosons arises from the
Higgs kinetic term in the Lagrangian. However, once we accept the SM
to be only an effective low-energy theory, higher-dimensional (and
hence non-renormalizable) terms are allowed.  
The most general $VVH$ vertex, consistent with Lorentz invariance
and current conservation\footnote{Terms not respecting current conservation
make vanishing contributions once a gauge boson couples to light fermions,
as at least one of them must in realistic experimental situations.}
can be written as
\begin{eqnarray}
\Gamma_{\mu\nu} &=& g_V\left[a_V \ g_{\mu\nu}+\frac{b_V}{m_V^2}
(k_{1\nu} k_{2\mu}
 - g_{\mu\nu}  \ k_{1} \cdot k_{2}) 
+\frac{\tilde b_V}{m_V^2} \ \epsilon_{\mu\nu\alpha\beta} \
k_1^{\alpha}  k_2^{\beta}\right]
\label{eq:coup}
\end{eqnarray}
where $k_i$ denote the momenta of the two $W$'s ($Z$'s). Here
\[
g_W^{SM} = e \, \cot\theta_W M_Z  \ , \qquad
g_Z^{SM}= 2 \, e M_Z/\sin2\theta_W, 
\]
$\theta_W$ being the weak-mixing angle and
$\epsilon_{\mu\nu\alpha\beta}$ the antisymmetric tensor with
$\epsilon_{0123}~=~1$.  Within the SM, $a_Z=a_W=1$ and $b_V=\tilde
b_V=0$ at tree level.  Anomalous parts may arise on account of higher
order contributions in a renormalizable theory~\cite{kniehl:NPB} or
from higher dimensional operators in an effective
theory~\cite{operator}.  While the imposition of ${\rm SU(2)_L \otimes
U(1)_Y}$ invariance relates the $WWH$ couplings to the $ZZH$ ones,
such an assumption restricts the nature of the physics beyond the SM.
Instead, we view them purely as phenomenological inputs and study their
effect on various final state observables in collider processes.

In general, each of these couplings can be complex, reflecting final
state interactions, or equivalently, absorptive parts of the loops
either within the SM or from some new physics beyond the SM.  However,
for each of the observables that we construct for the process $e^+ e^-
\to f \bar f H$, one overall phase can always be rotated away and
  we may choose that to be corresponding to either $a_Z$ or $a_W$. In
  our analysis, we choose $a_Z$ to be real and allow the others to be
  complex. Further we also assume $a_Z$ and $a_W$ to be close to there
  SM value i.e. $a_V = 1 + \Delta a_V$.

For a generic multi-doublet model, supersymmetric or otherwise,  
couplings of the neutral Higgs bosons to a pair of vector bosons ($V =
Z, W$) obey a sum rule~\cite{Gunion:1990kf,sumrule,Choudhury:2003ut}:
\begin{eqnarray}
 \sum_i a^2_{VVH_i} = 1.
\label{eq:sum}
\end{eqnarray}
Although $a_{VVH_i}$ for a given Higgs boson in different models, such
as MSSM, can be significantly smaller than the SM value, the presence
of higher $SU(2)_L$ multiplets or more complicated symmetry breaking
structures (such as those within higher-dimensional
theories)~\cite{Choudhury:2003ut} would lead to more complicated sum
rules.

The terms containing $a_V$ and $b_V$ in Eq.~\ref{eq:coup} constitute
the most general coupling of a $CP$-even Higgs boson with two vector
bosons whereas the $\tilde b_V$ term corresponds to the $CP$-odd
one. Simultaneous presence of both sets would indicate
$CP$-violation. A non-vanishing value for either $\Im(b_V)$ or
$\Im(\tilde b_V)$ destroys the hermiticity of the effective theory.

In the context of ${\rm SU(2)_L \otimes U(1)_Y}$ symmetry, the
couplings $b_V$ and $\widetilde b_V$ can be realized as first order
corrections arising from dimension-six operators such as $F_{\mu \nu}
F^{\mu \nu} \Phi^\dagger \Phi$ or $F_{\mu \nu}\tilde{F}^{\mu \nu}
\Phi^\dagger \Phi$ where $\Phi$ is the usual Higgs doublet, $F_{\mu
\nu}$ the field strength tensor and $\tilde{F}_{\mu \nu}$ its
dual~\cite{operator}. Of course, higher-order terms may also
contribute.  Equivalently, the relevant coupling constants may be
thought of as momentum dependent form factors.  However, for a theory
with a cut-off scale $\Lambda$ large compared to the energy scale at
which the scattering experiment is to be performed, the form-factor
behaviour would be very weak and hence can be neglected for our
study. Keeping in view the purported higher-order nature of the
anomalous couplings, we shall retain only terms up to the linear order
in all our expressions.

It is worthwhile to note here that, while our Eq.~\ref{eq:coup} is the most 
general expression for the $VVH$ vertex, consistent with Lorentz invariance 
and current conservation,
the process under consideration, namely $e^+ e^- \rightarrow H f \bar f$, 
can, in fact, receive anomalous/non-SM contributions from additional possible 
operators in an effective theory.  Examples include contact
interactions such as the dimension-6 $ff V H$ operator
~\cite{Buchmuller:1985jz}
\[
  \frac{\lambda_{\cal F}}{\Lambda^2} \;
  (\Phi^\dagger D_\mu \Phi) \; (\bar {\cal F} \gamma^\mu {\cal F})
\]
where ${\cal F} (\ni f/e)$ denotes a $SU(2)$ multiplet.
Even dimension-8 terms such as
\[
  \frac{g_{e f}}{\Lambda^4} \;
   (\Phi^\dagger \Phi) \; (\bar e \gamma^\mu e) \; (\bar f \gamma_\mu f)
\]
could contribute. The second term can arise from ultraviolet-complete
theories, such as a theory with a $Z'$ and accommodating a $Z' Z' H$
vertex in the limit of a very heavy $Z'$. The first one, on the other
hand, would require a $Z' Z H$ vertex as well.  Other constructions,
such as theories living in higher dimensions, could also lead to such
terms in an appropriate approximation~\cite{Choudhury:2003ut}.
While the $g_{e f}$ terms can be neglected in an effective theory
approach, the $\lambda_{\cal F}$ terms obviously have to be included
in the most general analysis of the process 
$e^+e^- \rightarrow f \bar f H$ 
\cite{Grzadkowski:1995hi,Kilian:1996wu,Rao:2006hn}.
Luckily, the contributions of such terms, arising say from a $Z'$ exchange,
to the $e^+ e^- \rightarrow f \bar f H$ amplitude, have the same structure
as that due to some of the terms in our anomalous vertex (Eq.~\ref{eq:coup}), 
as long as $\lambda_{\cal F}$  are flavour universal. 
For a generic theory---say, 
with a $Z'$ whose couplings to fermions are {\it not} flavour universal---the 
two contributions
may be distinguished from each other by a comparison of possible differences
in different channels. In the present work we desist from doing so
and thus implicitly assume a flavour universality of the underlying
UV-completion (say, the  $Z'$ couplings). The only remaining dimension-6
operator that is relevant to the given process is of the form 
$ \left( \bar {\ell}  \,  D_\mu \, {\rm e}  \right) \; (D^\mu \phi) $, 
where $\ell$ and ${\rm e}$ are fermionic $SU(2)$ doublet and singlet 
respectively. However, owing to a different chirality structure, it does not
interfere with the SM amplitude for a massless fermion 
and hence the corresponding contribution is highly
suppressed ($\sim~\Lambda^{-4}$).

%
Various terms in the effective $VVH$ vertex have definite properties 
under the discrete transformations $CP$ and $\tilde T$, where $\tilde T$
stands for the pseudo-time reversal transformation, one which reverses
particle momenta and spins but does not interchange initial and final
states. Table~\ref{tab:coup} shows the behaviour under the transformations,
 $CP$ and $\tilde T$ of various operators in the
effective Lagrangian, involving different coefficients given in the
table.  

\begin{table}[!h]
\begin{center}
\begin{tabular}{c|ccccc|}
\cline{2-6}
\\[-2.5ex]
\multicolumn{1}{c|}{} & $a_V$  & $\Re(b_V)$ & $\Im(b_V)$ &
$\Re(\tilde b_V)$ & $\Im(\tilde b_V)$\\ 
\hline
\multicolumn{1}{|c|}{$CP$} & $+$  & $+$ & $+$& $-$ & $-$ \\
\multicolumn{1}{|c|}{$\tilde T$} & $+$  & $+$ & $-$& $-$ & $+$
\\
\hline
\end{tabular}
\end{center}
\vskip -0.4cm
\caption{\em Transformation properties of 
the various operators (identified by their coefficients) in the
effective Lagrangian.}
\label{tab:coup}
\end{table}

\section{Kinematics and cuts}
\label{sec:cuts}
In this analysis, we largely consider the case of ILC operating 
at a center of mass energy of 500 GeV
and focus on the
case of an intermediate mass Higgs boson ($2m_b \leq m_H \leq 140$
GeV), for which $H\to b\bar b$ is the dominant decay mode with a
branching fraction $\gsim 0.68$~\cite{Djouadi:1997yw}.  To be
specific, we choose the mass of the Higgs boson to be
120 GeV and the $b$-tagging efficiency to be 0.7.

To be detectable, each of the final state particles in the process
$e^+e^- \to f \bar f H (b\bar b)$, must have a minimum energy and a
minimum angular deviation from the beam pipe.  Moreover, to be
recognized as different entities, they need to be well
separated. On the other hand, if the final state contains 
neutrinos, then the event must be characterized  by a
minimum missing transverse momentum. Quantitatively, the requirements
are
\begin{eqnarray}
\begin{array}{rclcl}
E_f &\ge&  10 \,\mbox{GeV} &&
\mbox{for each visible outgoing fermion}\\[0.2cm]
5^\circ~ \le ~ \theta_0 &\le& 175^\circ& &
\mbox{for each visible outgoing fermion}\\[0.2cm]
p_T^{\mbox {miss}} & \geq & 15\, \mbox{GeV} & &
\mbox{for events with} ~\nu'\mbox{s} \\[0.2cm]
\Delta R_{j j} & \geq & 0.7 & &
\mbox{for each pair of jets}\\[0.2cm]
\Delta R_{\ell\ell} & \geq & 0.2 & &
\mbox{for each pair of charged leptons}\\[0.2cm]
\Delta R_{l j}&\ge& 0.4&&
\mbox{for jet-lepton isolation }.
\end{array}
\label{eq:cuts}
\end{eqnarray}
Here $(\Delta R)^2 \equiv (\Delta \phi)^2 + (\Delta \eta)^2 $,
 $\Delta \phi $ and $\Delta \eta$ being the separation between the two
entities in azimuthal angle and rapidity respectively.

In addition, cuts may be imposed on the
invariant mass of the $f\bar f$ system to enhance(suppress) the
contributions coming from $s$-channel $Z$-exchange process, namely
\begin{equation}
\begin{array}{rcl}
R1 &\equiv& \left| m_{f\bar f} - M_Z \right| \leq 5 \, \Gamma_Z \hspace{0.5cm}
\mbox{ $\Longrightarrow$ select \ $Z$-pole} \ ,\\[1ex]
R2 &\equiv& \left| m_{f\bar f} - M_Z \right| \geq 5 \, \Gamma_Z \hspace{0.5cm}
\mbox{ $\Longrightarrow$ de-select \ $Z$-pole} \ , 
\end{array}
 \label{cuts:Z}
\end{equation}
where $\Gamma_Z$ is the width of the $Z$ boson. 
For the $\nu\bar\nu H$ 
final state, the same goal may be reached instead 
by demanding
\begin{equation}
\begin{array}{rcl}
 R1^\prime &\equiv & E^-_H \leq E_H \leq E^+_H,  \nonumber
\\[1ex]
R2^\prime &\equiv & E_H < E^-_H \ \mbox{or} \ E_H > E^+_H,
\end{array}
\label{cuts:EH}
\end{equation}
where $E_H^\pm = (s + m_H^2 - (m_Z\mp5\Gamma_Z)^2) / 
(2\sqrt{s})$. 

As described earlier, given that 
the anomalous couplings (${\cal B}_i$) correspond to operators  
that are notionally suppressed by some large scale, 
we  need retain terms only upto the
linear order. So, any observable ${\cal O}$ may be expressed as 
\begin{eqnarray}
{\cal O}(\{{\cal B}_i\}) &=& {\cal O}_{\rm SM} + 
              \sum \ O_i \ {\cal B}_i.\nonumber
\end{eqnarray}
The possible sensitivity of these observables to ${\cal B}_i$, at a
given degree of statistical significance $f$, can be obtained by
demanding that $|{\cal O}(\{{\cal B}_i\})-{\cal O}_{\rm SM}|\le f
\ \Delta{\cal O}$.  Here ${\cal O}(\{0\})={\cal O}_{\rm SM} $ is the
SM value of ${\cal O}$ and $\Delta{\cal O}$ is the fluctuation in the
measurement of ${\cal O}$, obtained by adding statistical and
systematic errors in quadrature.  For example, for ${\cal O}$ being
the total cross section ($\sigma$) or some asymmetry ($A$), we have
\begin{equation}
\begin{array}{rcl}
(\Delta\sigma)^2 &=& \sigma /{\cal L} + \epsilon^2 \, \sigma^2,
\\[1ex]
(\Delta A)^2 &=& \displaystyle
\frac{1-A^2}{\sigma\, {\cal L}} +
\frac{\epsilon^2}{2}(1-
A^2)^2.
\end{array}
\label{fluctuation}
\end{equation}
Here, $\cal L$ is the integrated luminosity of the
$e^+ e^-$ collider and  $\epsilon$ is the fractional systematic error
in cross section measurements. Throughout our analysis, we shall take $f=3$ and $\epsilon = 0.01$. In the approximation 
of retaining only terms upto linear order in anomalous couplings, 
the SM values for  $\sigma$ and $A$ are to be used in Eqs.~\ref{fluctuation}. 
\comment
{There has been some debate as to whether the fluctuations should be 
defined in terms of the SM expectations or the observed values. However, 
in our case, the two approaches give virtually the same numerical result.}

It may be noted here that in certain cases, such as when 
\[
\sigma_{\rm SM} > \frac{1}{{\cal L} \epsilon^2}
\]
the fluctuations are dominated by the systematic error and thus 
\[
(\Delta\sigma)^2 \approx  \sigma^2\epsilon^2 \ .
\label{syst_dom} 
\]
For example, the cross sections with the $R2^{\prime}$-cut for neutrino
final state that are used to constrain $WWH$ couplings in
Sec.~\ref{sec:wwh} satisfy this condition and hence the bounds on
these couplings are dominated by systematic errors.

\section{\label{sec:zzh} Anomalous $ZZH$ Couplings} %
The analysis of
Ref.~\cite{Biswal:2005fh} for the case of unpolarised beams had
revealed that various asymmetries probing the $ZZH$ anomalous
couplings were in fact proportional to $(l_{f}^2 - r_{f}^2)$ where
$l_{f} \, (r_{f})$ denote the coupling of a $Z$ boson to a
left-(right-)handed fermion. Use of polarised beams or detection of a
$\tau$ with a specific polarisation in the final state can then avoid
the cancellation between these two terms and may lead to an
enhancement in sensitivity.  In this section, we analyse various
observables that can be constructed with the use of these two
quantities.


\subsection{\label{sec:obs}Observables} 

Starting from various kinematical quantities $C_i$ constructed as 
various combinations of the different particle momenta and
their spins, we define observables ${\cal O}_i$, as expectation values
of the signs of $C_i$, i.e.  ${\cal O}_{i} = \langle {\rm
sign}(C_{i})\rangle$ ($C_{i}$'s, $i \ne 1$).  Each of these observables
transform in a well-defined manner under $C$, $P$ and $\tilde T$, and
within the aforementioned linear approximation, may be used to probe
the contribution of a given operator(s) in the effective Lagrangian
with the same transformation properties\footnote{Henceforth, we shall
interchangeably use the terminology ``transformation properties of
anomalous couplings'' for those of the corresponding operator in the
effective Lagrangian.}.  In fact, the observables listed here are the
same ones as considered in Ref.~\cite{Biswal:2005fh}, but here we use
them for a specific polarisation of initial beams and final state
$\tau$'s.  Needless to say, we concentrate on the cases where use of
polarisation affords a distinct gain in sensitivity. Table \ref{tab:finalcorr}
lists some of these
observables (cross sections and various asymmetries),
their transformation properties and the anomalous coupling they may
constrain; and, below, we give a description of the same:
\begin{table}[!h]
\begin{center}
\begin{tabular}{|r|l|ccccc|c|c|c|}
\hline
\hline
ID&${\cal C}_i$ & {$C$} & {$P$} & {$CP$} & {$\tilde T$} &
{$CP \tilde T$} &
$\begin{array}{c}
\mbox{Observa-}\cr
\mbox{ble}( {\cal O}_{i})
\end{array}$
& \mbox{Coupling} \\
\hline
\hline
&&&&&&&&\\[-3mm]
1 &$1 $ & {$+$} & {$+$} & {$+$} & {$+$}
& {$+$} &$\sigma$& {$a_z,\Re(b_z)$}  \\
\hline
&&&&&&&&\\[-3mm]
2&{$\vec{P}_{e} \cdot \vec{p}_H$} & {$-$} & {$+$} & {$-$}
& {$+$} & {$-$} &$A_{\rm FB}$& $\Im(\tilde b_z)$ \\
\hline
&&&&&&&&\\[-3mm]
3&{$(\vec{P_e}\times \vec{p}_H) \cdot \vec{P}_{f}$}
& {$+$} & {$-$} & {$-$} & {$-$} & {$+$} &$A_{\rm UD}$& $\Re(\tilde b_z)$  \\
\hline
&&&&&&&&\\[-3mm]
4&{$[\vec{P_e} \cdot \vec{p}_H] *
[(\vec{P_e}\times \vec{p}_H) \cdot \vec{P}_{f}] $}
& {$-$} & {$-$} & {$+$} & {$-$} & {$-$} &$A_{\rm comb}$& $\Im(b_z)$ \\
\hline
&&&&&&&&\\[-3mm]
5&{$[\vec{P_e} \cdot \vec{p}_{f}] *
[(\vec{P_e}\times \vec{p}_H) \cdot \vec{P}_{f}] $}
& {$\otimes$} & {$-$} & {$\otimes$} & {$-$} & {$\otimes$}
&$A{^{\prime}}_{\rm{comb}}$& $\Im(b_z), \Re(\tilde b_z)$  \\
\hline
\hline
\end{tabular}
\caption{ \label{tab:finalcorr}
{\em Various possible ${\cal C}_i$'s, their
transformation properties, the associated observables ${\cal O}_i$ 
and the anomalous couplings on which they provide information.
The symbol $\otimes$ indicates the lack of a definitive transformation 
property. Here, $\vec{P}_e \equiv \vec{p}_{e^-} -
\vec{p}_{e^+}$, $\vec{P}_f \equiv \vec{p}_{f} - \vec{p}_{\bar f}$
 and $\vec{p}_H$ is the momentum of Higgs boson (to be deduced from
the measurement of its decay products).}}
\end{center}
\end{table}

\begin{description}
\item[1. ${\cal O}_1$] is nothing but the total cross section as
obtained with a specific choice of polarisation 
for the initial beams and/or that for a final state $\tau$. 
As we retain  contributions  only
upto the lowest non-trivial order in the anomalous couplings (keeping
in view the higher dimensional nature of their origin), the 
differential cross section can be expressed as
%
\bea
d\sigma &=& \sum_{V=Z,W}[(1+ 2\; \Re(\Delta a_V)) d\sigma_{0V}
+ \Im(\Delta a_V) d{\sigma}^{\prime}_{0V} 
+ \Re(b_V) d\sigma_{1V} \nn \\ 
&& \qquad \qquad +~\Re(\tilde b_V) d\sigma_{2V} 
+ \Im( b_V) d\sigma_{3V} 
+ \Im(\tilde b_V) d\sigma_{4V}].
\label{def-cross}
\eea
where, as in Ref. \cite{Biswal:2005fh}, we have assumed that the Higgs 
is  SM-like  and hence
\beq
    a_V \equiv 1 + \Delta a_V
\eeq
is close to its SM value. 
As stated before we choose $a_Z$ to be real, hence 
$\Im(\Delta a_Z)$ = 0 and we denote $\Re(\Delta a_Z)$ = $\Delta a_Z$.  

\item
[2. ${\cal O}_2$]
is simply the forward-backward(FB) asymmetry  with respect to 
polar angle of the Higgs boson, namely
\begin{eqnarray}
A_{FB} (\cos\theta_H) &=& 
\frac{\sigma (\cos\theta_H>0) - \sigma (\cos\theta_H < 
0)}{\sigma (\cos\theta_H > 0) + \sigma (\cos\theta_H<0)} 
\label{eq:Afb}
\end{eqnarray}
Since ${\cal C}_2 \equiv \vec{P}_{e} \cdot \vec{p}_H$ is 
odd under $CP$ and even under $\tilde T$ transformation, this 
observable would thus be proportional to $\Im(\tilde b_z)$.
\item
[3. ${\cal O}_{3}$] is the 
up-down(UD) asymmetry defined in terms of the momentum of the 
final state fermion $f$ with respect to 
the $H$-production plane: 
\begin{eqnarray}
A_{UD} &=& \frac{\sigma (\sin\phi>0) - \sigma (\sin\phi<0)}
{\sigma (\sin\phi>0) + \sigma (\sin\phi<0)}\ .
\end{eqnarray}
As ${\cal C}_3 \equiv (\vec{P_e}\times \vec{p}_H) \cdot \vec{P}_{f}$
is odd under both
$CP$ and $\tilde T$, one may use this to
probe $\Re(\tilde b_Z)$.

\item
[4. ${\cal C}_{4}$] $\equiv [\vec{P_e} \cdot \vec{p}_H] *
[(\vec{P_e}\times \vec{p}_H) \cdot \vec{P}_{f}] $ is 
even under $CP$ and odd under $\tilde T$ and thus expected 
to be sensitive to $\Im(b_Z)$. The corresponding observable 
${\cal O}_4$ is a particular combination of 
the polar and azimuthal asymmetries (designed  to increase sensitivity)
and is defined as
\beq
A_{comb}  =
\frac{\sigma_{FU} +\sigma_{BD} -\sigma_{FD} -\sigma_{BU} }
{\sigma_{FU} +\sigma_{BD} +\sigma_{FD} +\sigma_{BU} } \ ,
\eeq
where F, B, U and D refer to the restricted phase space as mentioned
above in ${\cal O}_{2}$ and ${\cal O}_{3}$. Thus $\sigma_{FU}$ refers
to the cross section with Higgs boson restricted to be produced in
forward hemi-sphere with respect to the initial state electron and the
final state fermion is produced above the $H$-production plane.  

\item
[5. ${\cal O}_{5}$] is yet another asymmetry derived from 
a combination of polar and azimuthal distributions and is given by:
\beq
{A'}_{comb} =
\frac{\sigma_{F'U}+\sigma_{B'D} -\sigma_{F'D} -\sigma_{B'U} }
{\sigma_{F'U} +\sigma_{B'D} +\sigma_{F'D} +\sigma_{B'U} }.
\eeq
Here F{$'$} (B$'$) refer to the production of $f$ in forward (backward)
 hemi-sphere with respect to the initial state $e^{-}$, 
whereas U and D are the same as defined before. 
This being both $P$- and $\tilde T$-odd and with no  specific $C$
transformation, can be used to probe both 
$\Im (b_Z)$ and $\Re (\tilde b_Z)$.
\end{description}
Note that the asymmetries ${\cal O}_{3}$, ${\cal O}_{4}$ and ${\cal O}_{5}$
require charge measurement of the final state particles and
hence events with light quarks in the 
final state can not be considered for these observables.

\comment{
All these observables corresponding to a
state $\cal P$ for the initial beams will appear with a superscript
$\cal P$, whereas for a definite helicity of the final state $\tau$ we
will denote them with a superscript $\lambda$, where $\lambda = L,R$
denotes left-handed and right-handed $\tau$ respectively). We shall
also consider combinations of these observables with different
polarisation states.}
%

\subsection{\label{sec:beamzzh} Effect of Beam Polarization}
For longitudinally polarized beams, the cross section can be written  
as
\vskip -.2cm
\begin{eqnarray}
\sigma({{\cal P}_{e^-},{\cal P}_{e^+}}) &=& \frac {1}{4}  \biggl[
(1+ {\cal P}_{e^-})(1+{\cal P}_{e^+}) \sigma_{RR} 
+(1+ {\cal P}_{e^-})(1-{\cal P}_{e^+}) \sigma_{RL}
 \biggr]
\nonumber \\
&+&  \frac {1}{4} \biggl[
(1- {\cal P}_{e^-})(1+{\cal P}_{e^+}) \sigma_{LR} 
+(1- {\cal P}_{e^-})(1-{\cal P}_{e^+}) \sigma_{LL}
\biggr] \ ,
\nonumber 
\end{eqnarray}
where $\sigma_{LR}$ corresponds to the case of 
the electron (positron) beams being 
completely left(right) polarized respectively, i.e. , ${\cal
P}_{e^-} = -1$, ${\cal P}_{e^+} = +1$. 
$\sigma_{RR},~ \sigma_{RL} ~\mbox{and}~
\sigma_{LL}$ are defined analogously.
While the ideal case of complete polarisation is
impossible to achieve, values of  80\%(60\%) polarisation for
$e^-(e^+)$ seem  possible at the ILC~\cite{RDR}. Taking these to be our
default values,  we denote:
\begin{equation}
\sigma^{-,+} = \sigma({{\cal P}_{e^-} = - 0.8, {\cal P}_{e^+} = 0.6})
\nonumber
\end{equation}
and similarly for other combinations, viz. 
$\sigma^{+,+},~ \sigma^{+,-}~ \mbox{and}~ \sigma^{-,-}$.  
We concentrate here on the observables
discussed in the Sec.~\ref{sec:obs} for specific polarization combinations.
We would find that polarization plays a crucial role in probing 
the CP-odd $ZZH$ couplings, while the improvement in sensitivity 
in the others is only marginal.

We quote all our results for an integrated luminosity of 500 fb$^{-1}$
and a degree of statistical significance $f$ = 3 assuming the
fractional systematic error to be 1\% i.e.  $\epsilon$ of
Eq.~\ref{fluctuation} to be $0.01$. Denoting the four possible 
polarization combinations by
\beq
\barr{rclcrcl}
a &:& (-,+) \, , & \qquad & b &:& (+,-) \, ,\\
c &:& (-,-) \, , & & d &:&(+,+) \, ,
\earr
\label{def-pol_states}
\eeq
we consider two possible ways of
dividing the luminosity amongst these runs, namely 
\beq
\barr{rclcl}
\mbox {option (i) : } & \qquad & {\cal L} 
    & = & 125 \, {\rm fb}^{-1}\quad \qquad \mbox{for each of } (a, b, c, d)
\\[1ex]
\mbox {option (ii) : } & \qquad & {\cal L} 
    & = & \Bigg\{ \barr{rl}
                  200 \, {\rm fb}^{-1} & \quad \mbox{for } (a, b) \\
                  50 \, {\rm fb}^{-1} & \quad \mbox{for } (c, d) 
                  \earr 
    \ .
\earr
   \label{lum_div}
\eeq
While option (i) is a straightforward choice, option (ii) is better
appreciated on realizing that polarization combinations $(c, d)$
suppress both SM $s$-channel processes as also $WW$-fusion. Thus, 
although these combinations maybe useful for certain physics beyond the 
SM, it is not certain whether such modes would find favour for 
generic search strategies.

\subsubsection{$\Delta a_Z$ and $\Re(b_Z)$} 
      \label{sec:az_rebz}

It is obvious that the contribution of $\Delta a_Z$ would be identical
in form to that within the SM and, thus, such a coupling can be probed
only through a deviation of the cross sections from the SM
expectations. As for $\Re(b_Z)$, the fact that this term too 
conserves each of $C$, $P$ and $\tilde T$, renders all asymmetries 
insensitive to it. Thus, this coupling too needs to be measured from 
cross sections alone. 

Clearly, just one measurement cannot resolve 
between these two couplings, and this problem was faced by the analysis 
of Ref.~\cite{Biswal:2005fh} as well. Presumably, with beam 
polarization being available, cross section measurement 
for a variety of polarization states would offer additional 
information. However, as far as the Bjorken process goes, 
this dependence is, understandably, trivial and identical for both of 
$\Delta a_Z$ and $\Re(b_Z)$. This is attested to by 
the first two rows of Table~\ref{cross_sec}, which in fact 
lists different anomalous contributions to cross sections  
corresponding to different initial state polarisation combinations
and different final states.  For $e^+ e^- \to e^+ e^- H$
though, two diagrams contribute, the usual $s$-channel one and 
an additional $t$-channel one ($ZZ$ fusion), with 
the polarization dependence of the latter being 
grossly different. To accentuate this, we may de-select the $Z$-pole 
(the $R2$-cut of Eq.~\ref{cuts:Z}) and the corresponding cross sections 
are displayed in Table~\ref{cross_sec}. 
For completely polarized $e^\pm$ beams, LR and RL are the $CP$-eigen 
states, whereas LL and RR states are $CP$-conjugate to each other. 
Hence $\sigma_{LL}$ and $\sigma_{RR}$ receive additional contribution 
from $\Im(\tilde b_Z)$ and this is reflected in 
Table~\ref{cross_sec}.  
In addition, this contribution is proportional to 
(${\cal P}_{e^-} + {\cal P}_{e^+}$) and thus would vanish 
if the average values of this quantity vanishes (as, for example, 
happens in the unpolarized case). More importantly, the $\Re(b_Z)$ 
contribution to the $ZZ$-fusion diagram has an opposite dependence on 
the product (${\cal P}_{e^-}  {\cal P}_{e^+}$) as compared to that 
of the $a_Z$ piece. This may be exploited to 
construct an appropriate observable, namely 
\beq
\barr{rcl}
 \mathcal O_A &=& \dis 1.3 (\mathcal O^{\prime}_{1a}+ \mathcal
O^{\prime}_{1b}) + (\mathcal O^{\prime}_{1c}+ \mathcal
O^{\prime}_{1d}) \\
&=& \dis \left[ 15.1 (1+ 2 \Delta a_z) + 0.038 \,\Re(b_Z) \right] \, {\rm fb}
\earr
\label{O1abcd}
\eeq
\begin{table}[!ht]
\begin{center}
\begin{tabular}{|c|l|c|c|c|}
\hline
{\bf Observable} &{\bf Description} & {\textbf{ ${ \sigma_{0Z}}$ }}& 
{\textbf{$ \sigma_{1Z}$}}
& {\textbf{ $ \sigma_{4Z}$ }}\\
\hline
$\mathcal O_{1a}$ & $\sigma^{-,+}(R1;\mu,q)$ & 23.9 & { $226 $} & { $0 $}\\
$\mathcal  O_{1b}$ &  $\sigma^{+,-}(R1;\mu,q)$
 & 17.9 & { $169 $} & { $0 $} \\
$\mathcal O^{\prime}_{1a}$ & {$\sigma^{-,+}(R2;e)$  } & 4.04  & { $1.46$ } & { $ 0.122 $} \\
$\mathcal O^{\prime}_{1b}$ &  $\sigma^{+,-}(R2;e)$   & 2.64 & { $ 0.715 $} & { $ -0.122 $}  \\
%
$\mathcal  O^{\prime}_{1c}$ & {$\sigma^{-,-}(R2;e)$  } & 3.29 & { $ - 1.34$ } & { $ 0.855 $} \\
$\mathcal  O^{\prime}_{1d}$ & {$\sigma^{+,+}(R2;e)$  } & 3.09 & { $ - 1.45$ } & { $- 0.855 $} \\
\hline
\end{tabular}
\end{center}
\caption{{\em Various anomalous contributions (as defined in}
  Eq.~\ref{def-cross}{\em ) to the cross sections $\sigma(R1;\mu,q)$ (for
  $\mu^\pm$ and light quarks in the final state with $R1$-cut) and
  $\sigma(R2;e)$ (for $e^\pm$ in the final state with $R2$-cut) for the
  four polarisation states. The rates are in femtobarns, for $\sqrt{s}
  = 500$ GeV.}}
\label{cross_sec}
\end{table}
That the contribution of $\Im(\tilde b_Z)$ in Eq.~\ref{O1abcd}
vanishes identically is easy to understand. 
Even though the relative weights of the two terms in Eq.~\ref{O1abcd}
can be tuned to reduce 
the coefficient of $\Re(b_Z)$ further, it is not really necessary, 
since both $\Delta a_Z$ and $\Re(b_Z)$ are expected to
arise, say at one-loop, and hence would have similar order of magnitude.  
The large difference in the relative weights renders $\Re(b_Z) $
almost irrelevant, making it plausible to constrain $ \Delta a_z$
independent of $\Re(b_Z)$. A lack of deviation of $ \mathcal O_A $
from its SM value would give a $3\sigma$ level  limit on $
\Delta a_z$ of the form 
\beq
|\Delta a_Z | \leq  \left\{
  \barr{rcl}
     0.038 & \quad & {\rm for \,\,option \,(i) }
     \\[0.5ex]
     0.043 & &  {\rm for \,\,option \,(ii)}.
  \earr
  \right.
\label{lim_az}
\eeq
The smallness of the difference in the two limits 
is but a consequence of the 
larger error bars resulting from smaller luminosities assigned to 
polarization combinations $(c,d$). 
Note that the limits are only marginally different from 
that obtained with unpolarised beams with ${\cal L} = 500\, {\rm
fb}^{-1}$, namely of 
$|\Delta a_Z | \lsim 0.04$~\cite{Biswal:2005fh}\footnote{Note 
that the results quoted here for the unpolarised beams
    differ from those of the Ref.~\cite{Biswal:2005fh} because of the
    difference in the value of branching fraction of $H \to b\, \bar
    b$ decay mode used there. We have used the value of branching
    fraction to be 0.68 whereas in Ref.~\cite{Biswal:2005fh} it was
    taken to be 0.9}.
More importantly, though, unlike the bound of Ref.~\cite{Biswal:2005fh}, 
the constraint of Eq.~\ref{lim_az}
is independent of $\Re(b_Z)$.

\begin{figure}[!h]
\begin{center}
\epsfig{file=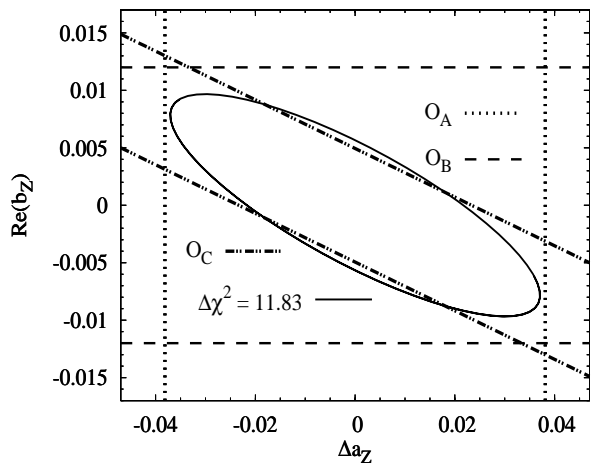,width=8.5cm,height=6.5cm}
\caption{\label{fig:az-rbz}{\em The regions in the $\Delta
    a_Z-\Re(b_Z)$ plane consistent with $3\sigma$ variations 
    in the observables $ \mathcal O_{A, B, C}$ respectively. 
    The region enclosed by all the three sets reflect the 
    overall constraints. 
The ellipse represents the region corresponding to $\Delta \chi^2 
= 11.83$ obtained using all polarised cross sections listed in 
Table~\ref{cross_sec}. An integrated
    luminosity of 125 fb$^{-1}$ for each of the polarisation state
    i.e. option (i) of Eq.~\ref{lum_div} has been used. The limits 
    for option (ii) are very similar.}}
\end{center}
\end{figure}
A different linear combination of the same observables, namely, 
\beq
\barr{rcl}
 \mathcal O_{\rm B} &=& \mathcal O_{1a} + \mathcal O_{1b} -6.6 \mathcal
 (O^{\prime}_{1c}+ \mathcal O^{\prime}_{1d}) 
\\[0.5ex] 
&=& \left[ -0.31( 1+ 2 \Delta  a_Z) + 413 \Re(b_Z) \right] \, {\rm fb}
\earr
\eeq
is equally useful as this enhances the $\Re(b_Z)$ contribution, while 
essentially getting rid of $\Delta  a_Z$. This leads to 
\beq
 |\Re(b_Z)| \leq \left\{
   \barr{rcl}
     0.012  & \qquad & {\rm for \,\,option\,\, (i) } 
 \\[0.5ex] 
     0.018 & &  {\rm for \,\,option\,\, (ii) } 
    \earr
   \right.
\label{lim_rbz_mix}
\eeq
virtually independent of $\Delta  a_Z$. Finally, using the 
information from the $R1$-cut alone, we have 
\beq
\barr{rcl}
  \mathcal O_C &=& \mathcal O_{1a} + \mathcal O_{1b}
\\[0.5ex]  
&=&  \left[ 41.8 (1+ 2 \Delta a_z) + 395 \,\Re(b_Z) \right] \, {\rm fb}
\earr
\label{O1aPlO1b}
\eeq
leading to a correlated constraint in the $\Delta  a_Z$--$\Re(b_Z)$
plane as displayed in Fig.~\ref{fig:az-rbz}. 
Of the six cross sections of Table~\ref{cross_sec}, one may be used 
to eliminate $\Im(\tilde{b}_{Z})$, leaving behind five constraints 
in this plane.  
Note that we have already used three linearly independent
combinations.  
We may, nonetheless, use all five to construct a $\chi^2$-test.
The resultant $3 \sigma$ ellipse (corresponding to $\Delta \chi^2 = 11.83$)
is also displayed in Fig.~\ref{fig:az-rbz}. That this ellipse protrudes
slightly beyond the set of straight lines is not surprising, for 
the latter 
denote the $3 \sigma$ constraint on a particular combination of the two 
variables (with complete disregard for the orthogonal combination), while
the ellipse gives the corresponding bound on the plane. Furthermore,
the size and the shape of the ellipse demonstrates that the three 
combinations identified above represent the strongest constraints 
with very little role played by the remaining two.

Finally, recollect that, for completely polarized $e^\pm$ beams, 
LL and RR states are $CP$-conjugate to each other. Hence 
the difference of $\sigma_{LL}$ and $\sigma_{RR}$ can be used to 
probe $CP$-odd couplings. Thus, using 
\beq
\mathcal O_D \equiv  \mathcal O^{\prime}_{1c} - \mathcal O^{\prime}_{1d}
= \left[ 0.2 \, (1 + 2 \Delta a_z) + \,
0.11 \,\Re(b_Z) + \, 1.71 \, \Im(\tilde{b}_{Z}) \right] \, {\rm fb}
\ ,
\label{Op1cd}
\eeq
one obtains
\bea
|\Im(\tilde{b}_{Z})| &\leq& 0.4 \quad {\rm for \,\, {\cal L} = 125\,  fb^{-1}}.
\label{lim-imtbz2}
\eea
However, a better constraint can be obtained on this coupling 
with the use of FB-asymmetry with respect to polar angle of 
Higgs boson and a discussion of this follows.

\subsubsection{$\Re(\tilde b_Z)$ and  $\Im(\tilde b_Z)$}

Next we focus on the role of beam polarisation in
exploring both the real and imaginary parts of $\tilde{b}_{Z}$. As discussed 
earlier, the independent experimental probes for these 
couplings, namely the asymmetries $A_{\rm FB}$ and $A_{\rm UD}$, are 
proportional to the quantity $(l_e^2 -r_e^2)$ 
for the case of unpolarised beams~\cite{Biswal:2005fh}.  For maximally 
polarized
beams, on the other hand, 
the squared matrix element is either proportional to $\resq$ or
to $\lesq$ and hence the suppression factor is not so severe even for 
moderate polarization.
Since $\lesq > \resq$, the cross sections are somewhat larger for the 
polarization combination $a \equiv (-, +)$, and hence the 
corresponding constraints would turn out to be a  little stronger.

Imposing the various kinematical cuts of Eq.~\ref{eq:cuts}, along-with
the $R1$-cut of Eq.~\ref{cuts:Z} to select $Z$-pole contribution, 
the forward-backward (referring to the Higgs polar angle) 
asymmetry for different final states and with different
polarisation states of the initial beams, can be written, keeping terms
up to linear order in anomalous couplings, as 
\begin{equation}
\begin{array}{rcl}
A^{-,+}_{FB}(R1-\rm{cut}) &=&
\left\{
\begin{array}{lcl}
\displaystyle
\frac{ 0.174 \ \Re(\tilde b_Z) - 6.14 \ \Im(\tilde b_Z)}{1.48 }
&  & (e^+ e^- H)\\[1.5ex]
\displaystyle
\frac{- 6.07 \ \Im(\tilde b_Z)}{1.46 }
&  & (\mu^+ \mu^- H) \\[1.5ex]
\displaystyle
\frac{- 92.8 \ \Im(\tilde b_Z)}{22.4} &  &
(q \bar q H)
\end{array}
\right.
\end{array}
\label{eq:asym_FB}
\end{equation}
and
\begin{equation}
\begin{array}{rcl}
A^{+,-}_{FB}(R1-\rm{cut}) &=&
\left\{
\begin{array}{lcl}
\displaystyle
\frac{ - 0.0911 \ \Re(\tilde b_Z) + 4.43 \ \Im(\tilde b_Z)}{1.11 }
&  & (e^+ e^- H)\\[1.5ex]
\displaystyle
\frac{ 4.4 \ \Im(\tilde b_Z)}{1.09 }
&  & (\mu^+ \mu^- H) \\[1.5ex]
\displaystyle
\frac{67.2 \ \Im(\tilde b_Z)}{16.8} &  &
(q \bar q H)
\end{array}
\right.
\end{array}
\label{eq:asym_FB2}
\end{equation} 
Here, each of the numerical factors denote cross-sections in
femtobarns with the denominator being the SM cross
section. Understandably, the FB asymmetry is identical for the case of
the $Z$ going into muons or a $q \bar q$ pair, and thus the two
channels can be added up to obtain the total sensitivity. As for the
$e^+ e^- H$ channel, the coupling $\Re(\tilde b_Z)$ makes an
appearance on account of the interference of the $t$-channel diagram
with the absorptive part of the $s$-channel SM one.

\begin{figure}[!ht]
\begin{center}
\epsfig{file=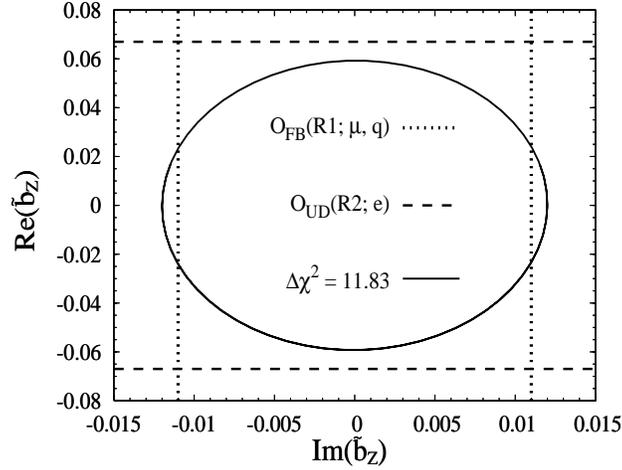,width=8.5cm,height=6.5cm}
\caption{\label{fig:imtbz-retbz3}{\em The regions in the
    $\Im(\tilde{b}_{Z})$--$\Re(\tilde{b}_{Z})$ plane
consistent with $3\sigma$ variations
    in the observables ${\cal O}_{FB}(R1; \mu, q)$ and
${\cal O}_{UD}(R2; e)$. The vertical and
horizontal lines are for respective variation in them. 
The ellipse represents the region corresponding to a $3 \sigma$ 
bound in the plane ($\Delta \chi^2
= 11.83$) obtained using all asymmetries listed in
Eqs.~\ref{eq:asym_FB},~\ref{eq:asym_FB2},~\ref{eq:asym_UD},
~\ref{eq:asym_UD2},~\ref{eq:asym_UD_r1r2}. Option (i) of luminosity 
division has been assumed. 
}}
\end{center}
\end{figure}

The FB-asymmetry with final state $\mu$'s and q's get 
contribution only from $\Im(\tilde{b}_{Z})$. Hence we 
define the following observable
\bea
{\cal O}_{FB}(R1; \mu, q) &=& 
A^{-,+}_{FB}(R1; \mu) \, + A^{-,+}_{FB}(R1; q) 
- \, A^{+,-}_{FB}(R1; \mu) \, - A^{+,-}_{FB}(R1; q) \nn \\
&=& -\, 16.3 \, \Im(\tilde{b}_{Z}) 
\label{eq:FB-asym1} 
\eea
which leads to 
\beq
|\Im(\tilde{b}_{Z})| \leq \left\{
\barr{lcl}
 0.011 & \quad  & {\rm for \,\,option \,(i) }\nn \\
 0.0089 & & {\rm for \,\,option \,(ii)}.
  \earr
  \right.
\label{lim_imtbz}
\eeq
In Fig.~\ref{fig:imtbz-retbz3}, the vertical lines 
represent the above bounds for option (i).

Similarly,  the up-down asymmetries, 
 with respect to azimuthal angle of final state fermions, are given by
\begin{equation}
\begin{array}{rcl}
A^{-,+}_{UD}(R1-\rm{cut}) &=&
\left\{
\begin{array}{lcl}
\displaystyle
\frac{-1.43 \ \Re(\tilde b_Z) - 0.286 \ \Im(\tilde b_Z)}{1.48 }
&  & (e^+ e^- H)\\[1.5ex]
\displaystyle
\frac{- 1.49 \ \Re(\tilde b_Z)}{1.46 }
&  & (\mu^+ \mu^- H) \\[1.5ex]
\end{array}
\right.
\end{array}
\label{eq:asym_UD}
\end{equation}
and
\begin{equation}
\begin{array}{rcl}
A^{+,-}_{UD}(R1-\rm{cut}) &=&
\left\{
\begin{array}{lcl}
\displaystyle
\frac{1.12 \ \Re(\tilde b_Z) - 0.161 \ \Im(\tilde b_Z)}{1.11}
&  & (e^+ e^- H)\\[1.5ex]
\displaystyle
\frac{1.08 \ \Re(\tilde b_Z)}{1.09}
&  & (\mu^+ \mu^- H) \\[1.5ex]
\end{array}
\right.
\end{array}
\label{eq:asym_UD2}
\end{equation}
Since the determination of this asymmetry requires charge measurement of the
final state particles, we do not consider it for quarks in
the final states. Once again, $\Im(\tilde b_Z)$ makes an appearance
for the $e^+ e^- H$ case on account of the aforementioned
interference. Combining both polarisation states $(-,+)$ and $(+,-)$
for final state muons we construct a observable, namely, 
\beq
{\cal O}_{UD}(R1; \mu) \equiv 
A^{-,+}_{UD}(R1; \mu)  
- \, A^{+,-}_{UD}(R1; \mu) \nn \\ 
= -\, 2.01 \, \Re(\tilde{b}_{Z})  \ ,
\label{eq:UD-asym1} 
\eeq
we may constrain 
\beq
|\Re(\tilde{b}_{Z})| \leq  \left\{
\barr{lcl}
 0.17 & \quad & {\rm for \,\,option \,(i) }\nn \\
0.13 & &  {\rm for \,\,option \,(ii)}.
\earr
\right.
\label{lim_retbz1}
\eeq

Clearly, one expects a nontrivial effect of the beam poalrisation only
for asymmetries constructed with the $R1$-cut. For the sake of completeness,
we also present the up-down asymmetries with the 
$R2$-cut (de-selecting the $Z$-pole) for the $eeH$ final state, namely 
\begin{equation}
\begin{array}{rcl}
A^{-,+}_{UD}(R2; e) &=& \dis
\frac{4.3 \, \Re(\tilde b_Z) + 0.227 \, \Im(b_Z) }{4.04}, 
\\[2ex]
A^{+,-}_{UD}(R2; e) &=& \dis 
\frac{3 \ \Re(\tilde b_Z) - 0.227 \, \Im(b_Z) }{2.64},
\\[2ex]
A^{-,-}_{UD}(R2; e) &=& \dis
\frac{4.01 \, \Re(\tilde b_Z) + 1.59 \, \Im(b_Z) }{3.29}, 
\\[2ex]
A^{+,+}_{UD}(R2; e) &=& \displaystyle
\frac{3.82 \, \Re(\tilde b_Z) - 1.59 \, \Im(b_Z) }{3.09}. 
\end{array}
\label{eq:asym_UD_r1r2}
\end{equation}
With LL and RR initial states being $CP$-conjugate to each other, it is
understandable that $A_{UD}$ receives 
additional contribution\footnote{This is analogous to the appearance 
of $\Im(\tilde b_Z)$ in the total cross sections in Sec.\ref{sec:az_rebz}.}
 from $\Im(b_Z)$. Defining 
\beq
\barr{rcl}
{\cal O}_{UD}(R2; e) &=&
2 \, A^{-,+}_{UD}(R2; e) + \, A^{+,-}_{UD}(R2; e)   
+ \, A^{-,-}_{UD}(R2; e) + \, A^{+,+}_{UD}(R2; e) 
\\[1ex]
&=&  5.72 \, \Re(\tilde{b}_{Z}) - 0.005 \, \Im(b_Z)
\earr
\label{eq:UD-asym2}
\eeq
one may get rid of this contribution, and thereby constrain
\beq
|\Re(\tilde{b}_{Z})| \leq \left\{
\barr{lcl}
   0.067 & \quad & {\rm for \,\,option \,(i) }
   \\
   0.074 & &  {\rm for \,\,option \,(ii)}
  \earr
  \right.
\label{lim_retbz}
\eeq
at the  $3 \sigma$ level. 
The bounds for option (i) are represented by horizontal lines
in Fig.~\ref{fig:imtbz-retbz3}.
  
%
%
\begin{table}[!h]
\begin{center}
\begin{tabular}{||cc||c|c|c||c|c||}
\hline\hline
\multicolumn{5}{||c||}{Using Polarized Beams} &
\multicolumn{2}{|c||}{Unpolarized States} \\[2mm]
\hline
Coupling & &
\multicolumn{2}{|c|}{Limits}
&
$\begin{array}{c}
\mbox{Observable}~ \cr
\mbox{used}
\end{array}$
& Limits &
$\begin{array}{c}
\mbox{Observable}~ \cr
\mbox{used}
\end{array}$
\\
\cline{3-4}
&&
\mbox{Option(i)}
&
\mbox{Option(ii)}
&
&& \\
\hline
\hline
&&&&&&\\
$|\Re(\tilde b_{Z}|$
& $\leq$ & 0.067 & 0.074 &
$\begin{array}{l}
{\cal O}_{UD}(R2; e) \cr
\end{array}$
& 0.067 &
$\begin{array}{l}
A_{UD}(R2; e) \cr
\end{array}$\\[2mm]
\hline
\hline
&&&&&&\\
$|\Re(\tilde b_{Z})|$ & $\leq$ & 0.17 & 0.13 &
$\begin{array}{l}
{\cal O}_{UD}(R1; \mu) \cr
\end{array}$
& 0.91 &
$\begin{array}{l}
A_{UD}(R1; \mu) \cr
\end{array}$\\[2mm]
\hline
\hline
&&&&&&\\
$|\Im(\tilde b_{Z})|$ & $\leq$ & 0.011 & 0.0089 &
$\begin{array}{l}
{\cal O}_{FB}(R1; \mu, q) \cr
\end{array}$
& 0.064 &
$\begin{array}{l}
A_{FB}(R1; \mu, q) \cr
\end{array}$\\[2mm]
\hline
\hline
\end{tabular}
\vskip -0.2cm
\caption{\label{tab:beamlimit} {\em Limits on anomalous $ZZH$
    couplings from various observables at $3\sigma$ level 
    for both polarized and unpolarized beams. While 
    an integrated luminosity of  $500 \, {\rm fb}^{-1}$ is 
    assumed for the unpolarized run, for the polarized case 
    the same is divided according to the options of Eq.~\ref{lum_div}.
}}
\end{center}
\end{table}

It is instructive to compare the above sensitivities (see 
Table~\ref{tab:beamlimit}) 
with those possible with unpolarised 
beams~\cite{Biswal:2005fh}. As one
can see, for asymmetries with $R1$-cut, the enhancement of 
sensitivity to both the $CP$-odd couplings  
$\Re (\tilde b_Z)$ and $\Im (\tilde b_Z)$ on using 
longitudinally polarized beams  for option (ii)(option (i)) of 
Eqs.~\ref{lum_div}) is nearly by a factor of 7(5--6), a feat 
unachievable without polarisation.
This improvement is indeed due to the
circumvention of the vanishingly small vector coupling
of electron to the $Z$ boson.  Our results are compatible with the 
more stringent limits of Ref.~\cite{Han:2000mi}, when we remove the
effect of kinematical cuts as well as of the use of the $b \bar b$
final state and finite $b$--tagging efficiency, implemented in our
analysis. Of course, a similar enhancement for $\Re(\tilde b_{z})$ 
may also be possible even with unpolarised beams, 
when $A_{UD}$ for the $eeH$ final state is considered 
alongwith the $R2$-cut.
It is, nonetheless, interesting to have two different
measurements measure the same coupling to the same accuracy. 
Of course, this in addition to the big enhancement
in sensitivity to the $CP$-odd couplings
$\Re (\tilde b_Z)$ and $\Im (\tilde b_Z)$ as mentioned
above; this,  in fact, is not achievable with the $R1$-cut.

%
%
%
A further improvement is possible if 
all the asymmetries listed in Eqs.~(\ref{eq:asym_FB},\ref{eq:asym_FB2},
\ref{eq:asym_UD},\ref{eq:asym_UD2}) are considered alongwith the three 
independent combinations obtained from Eqs.~\ref{eq:asym_UD_r1r2} by 
eliminating $\Im(b_Z)$ therefrom. Once again, a $\chi^2$ 
can be constructed leading to a $3\sigma$ contour as displayed in 
Fig.~\ref{fig:imtbz-retbz3}.

And, finally, we use the remaining information available from 
Eqs.~\ref{eq:asym_UD_r1r2} to probe $\Im(b_Z)$. Defining
\beq
{\cal O}^{\prime}_{UD}(R2; e) \equiv
A^{-,-}_{UD}(R2; e) - \, A^{+,+}_{UD}(R2; e) \nn \\
=  - \, 0.017 \, \Re(\tilde{b}_{Z}) + 0.998 \, \Im(b_Z) \ ,
\label{eq:UD-asym3}
\eeq
we may impose 
\bea 
|\Im(b_Z)| &\leq & 0.22 \quad {\rm for} \,\, {\cal L} = 125\, {\rm fb}^{-1}. 
\label{lim-imbz-n}
\eea
This limit is only marginally different than the one obtained with 
unpolarised beams using the combined asymmetry $A_{comb}$ of
 Sec.~\ref{sec:obs} with $R1$-cut. We shall see in the next section 
that a better constraint may be obtained on this coupling from the 
combined asymmetry if the final state $\tau$ helicity is measured.

\subsection{\label{sec:finaltau} Use of Final State $\tau$
  Polarization with Unpolarised Beams.}
In this section we report on  the use of  selecting final states 
with  $\tau$'s  in a given helicity state. A similar idea was 
employed in the optimal variable analysis of Ref.~\cite{Hagiwara:2000tk}.
A detailed measurement of the decay pion energy 
distribution~\cite{Hagiwara:1989fn}
or a simpler measurement of the inclusive single pion 
spectrum~\cite{Roy:1991sf}, 
can yield information on $\tau$ polarisation. Discussions exist in literature
on how this can be utilized to sharpen search strategies for charged Higgs
boson~\cite{Hagiwara:1989fn,Roy:1991sf,Bullock:1991fd,Bullock:1992yt,
Raychaudhuri:1995kv,Raychaudhuri:1995cc,Bella:1994dk} or 
supersymmetric partners~\cite{guchaitroy}
and even for the measurement of SUSY parameters~\cite{nojiri,Godbole:2004mq}.
In a similar vein, one can also construct observables using the
final state $\tau$ polarisation to probe $ZZH$ couplings. 
We construct asymmetries for $\tau$ 
with definite polarisation which can be measured in simple counting experiments
and which can catch the essence of the above mentioned optimal observable 
analysis.

As discussed in Ref. \cite{Biswal:2005fh}, for unpolarized 
initial and final states, the
combined polar-azimuthal asymmetry $A_{\rm comb}$ is proportional to
($\lesq + \resq$)($\rfsq - \lfsq$) and the up-down asymmetry,
$A_{UD}(\phi)$ is proportional to ($\lesq - \resq$)($\rfsq -
\lfsq$). Thus, for leptonic final states, both these asymmetries 
suffer a suppression (this is particularly important for both  
$A_{UD}$ and $A_{\rm comb}$ which are impossible to measure 
with hadronic final states). Hence, the measurement of the 
final state $\tau$ polarisation 
would lead to an enhancement in these symmetries with a consequent
improvement in the sensitivity to both of the $\tilde T$-odd anomalous
couplings, namely $\Im (b_Z)$ and $\Re (\tilde b_Z)$.
Further, since $\ltsq > \rtsq$, one gets a slightly
higher gain in sensitivity with final state $\tau$ in a negative
helicity state.

To demonstrate this, we construct various asymmetries (listed in
Table~\ref{tab:finalcorr}) for both left- and right-handed $\tau$ in
the final state. These are the same as described in
Sec.~\ref{sec:obs} but defined for a particular helicity of $\tau$
rather than for a specific initial beam polarisation state (for this 
analysis we take the initial state to be unpolarised). 
Again, after imposing the kinematical cuts of
Eq.~\ref{eq:cuts} alongwith the $R1$-cut, these read
\beq
\begin{array}{rcl}
A^L_{UD} (R1; \tau) &=& \displaystyle
\frac{-0.527  \ \Re(\tilde b_Z)}{0.495},\\[1.7ex]
A^R_{UD} (R1; \tau) &=& \displaystyle
\frac{0.388  \ \Re(\tilde b_Z)}{0.365} ,\\[1.7ex]
A^L_{\rm comb} (R1; \tau) &=& \displaystyle
\frac{- 1.37 \; \Im(b_Z)}{0.495 },\\[1.7ex]
A^R_{\rm comb} (R1; \tau) &=& \displaystyle
\frac{ 1.01 \; \Im(b_Z)}{0.365 },\\[1.7ex]
A{^{\prime}}_{\rm comb}^L (R1; \tau)& =& \displaystyle
\frac{ 1.18 \; \Im(b_Z)~-~0.3~\Re (\tilde b_Z)}{0.495 },\\[1.7ex]
A{^{\prime}}_{\rm comb}^R (R1; \tau)& =& \displaystyle
\frac{ -0.868 \; \Im(b_Z)~-~0.221~\Re (\tilde b_Z)}{0.365 }.
\end{array}
\label{asymm_tau}
\eeq
where the numbers represent the partial (total) cross sections 
in femtobarns. As expected, the asymmetries are, generically,
in the opposite sense
for left- and right-handed $\tau$'s (and, hence, would have 
largely cancelled each other). Thus, being 
able to isolate $\tau$'s of a given helicity indeed can add to the
sensitivity and the slightly larger value of $l_{\tau}$ means that the
final state with negative helicity $\tau$'s will be even better.
Rather than effect a detailed analysis, we make a simplifying and 
{\em illustrative} assumption that final state $\tau^-$'s
in a specific (negative or positive) helicity state can be isolated 
with very high purity at the cost of an efficiency of 40\% or even 20\%. 
Under this assumption, we list, in
Table~\ref{tab:finaltauL}, the best limits that are possible for the
$\tilde T$-odd couplings $\Im (b_z)$ and $\Re (\tilde b_Z)$ using
asymmetries of Eq.~\ref{asymm_tau} for negative helicity
$\tau$'s. (The positive helicity $\tau$'s would lead to 
very similar (marginally weaker) bounds, and we omit these for 
ease of presentation.) 
Corresponding limits obtainable with the $\tau$ final state but
without using knowledge of $\tau$ helicity are also shown for
comparison.
\begin{table}[!h]
\begin{center}
\begin{tabular}{||cc||c|c|c||c|c||}
\cline{3-7}
\multicolumn{2}{c|}{}& \multicolumn{3}{c||}{Using Pol. of  final state $\tau^-$} &
\multicolumn{2}{|c||}{Unpolarised $\tau$'s} \\[2mm]
\hline
Coupling & &
\multicolumn{2}{c|}{Limits}
&
\mbox{Observable}
& Limits &
\mbox{Observable}
\\
\cline{3-4}
&&
\mbox{40\% eff.}
&
\mbox{20\% eff.}
&
&& \\
\hline
&&&&&&\\[-2mm]
$|\Im(b_{z})|$
& $\leq$ & 0.11 & 0.15 &
$\begin{array}{l}
A_{comb}^{L} \cr
\end{array}$
& 0.35 &
$\begin{array}{l}
A_{comb} \cr
\end{array}$\\
\hline
&&&&&&\\[-2mm]
$|\Re(\tilde b_{z})|$ & $\leq$ & 0.28 & 0.40 &
$\begin{array}{l}
A_{UD}^{L} \cr
\end{array}$
& 0.91 &
$\begin{array}{l}
A_{UD} \cr
\end{array}$\\
\hline
\end{tabular}
\vskip -0.2cm
\caption{\label{tab:finaltauL} {\em Limits on anomalous $ZZH$ couplings
from various observables at $3 \sigma$ level 
with negatively polarised and unpolarised $\tau$'s (i.e. without 
using knowledge of $\tau$ helicity),  with $R1$-cut.}}
\end{center}
\end{table}

It is clear from Table~\ref{tab:finaltauL} that the limits on 
$\Im (b_Z)$ and $\Re (\tilde b_Z)$ can be improved by the measurement 
of the final state $\tau$ polarization.  Preliminary results of  
our analysis were presented in Ref.~\cite{Biswal:2007vy}.
A similar observation had been made earlier in the
context of optimal observable analysis~\cite{Hagiwara:2000tk}. It is
to be noted here that the unpolarized measurements with $eeH$ final
state, with $R2$-cut gives a better sensitivity to $\Re (\tilde b_Z)$ as
discussed earlier.  It is, however, nice to have more than one
observable determining a given coupling.  We would like to stress
here that even with an efficiency of isolating events with a given 
$\tau^-$
helicity as low as 20\%, this method 
affords an increase in the sensitivity for $\Im(b_Z)$ by as much as a
factor $\sim 2$.

\begin{figure}[!h]
\centerline{
\epsfig{file=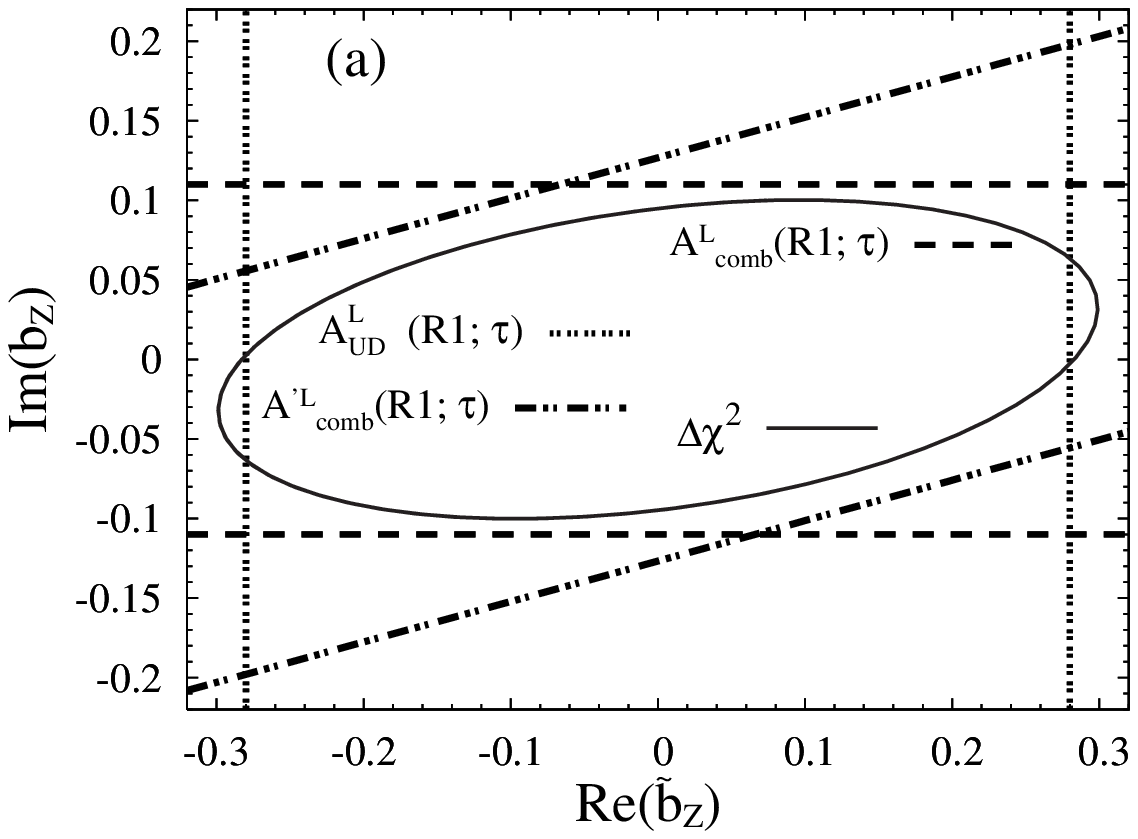,width=5.5cm,height=6.0cm}
\epsfig{file=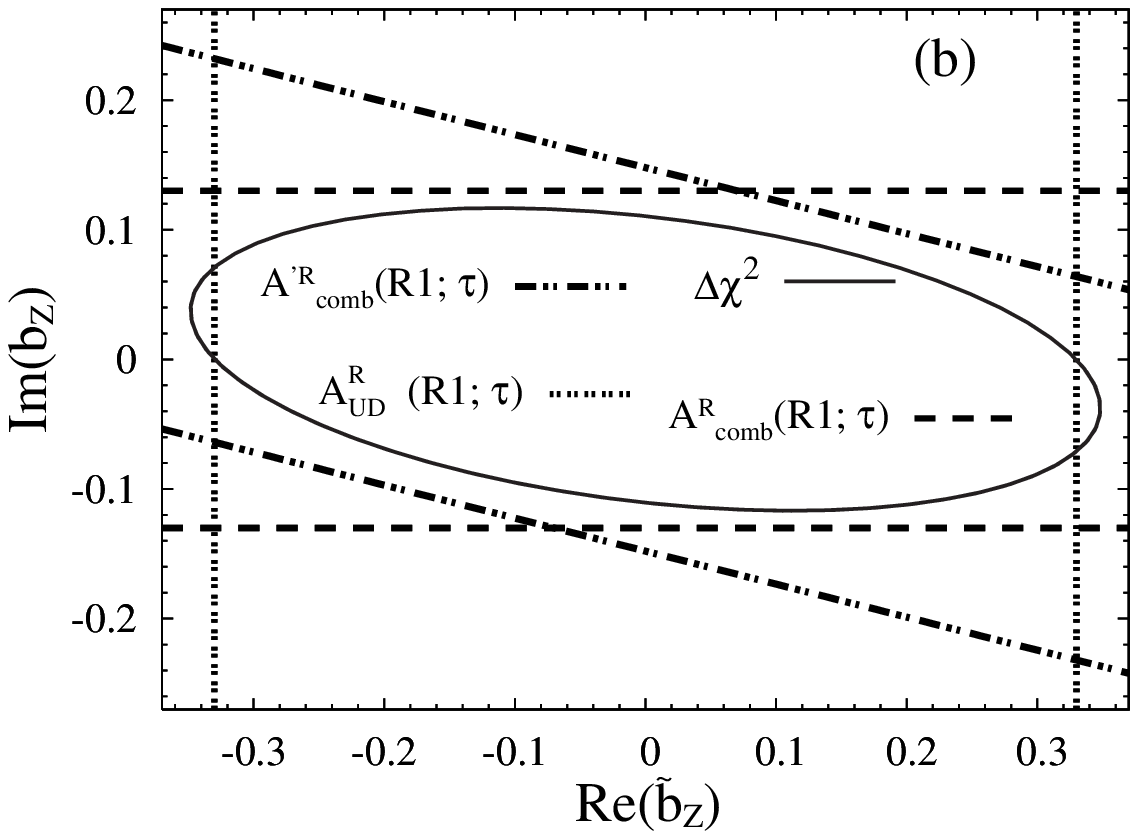,width=5.5cm,height=6.0cm}
\epsfig{file=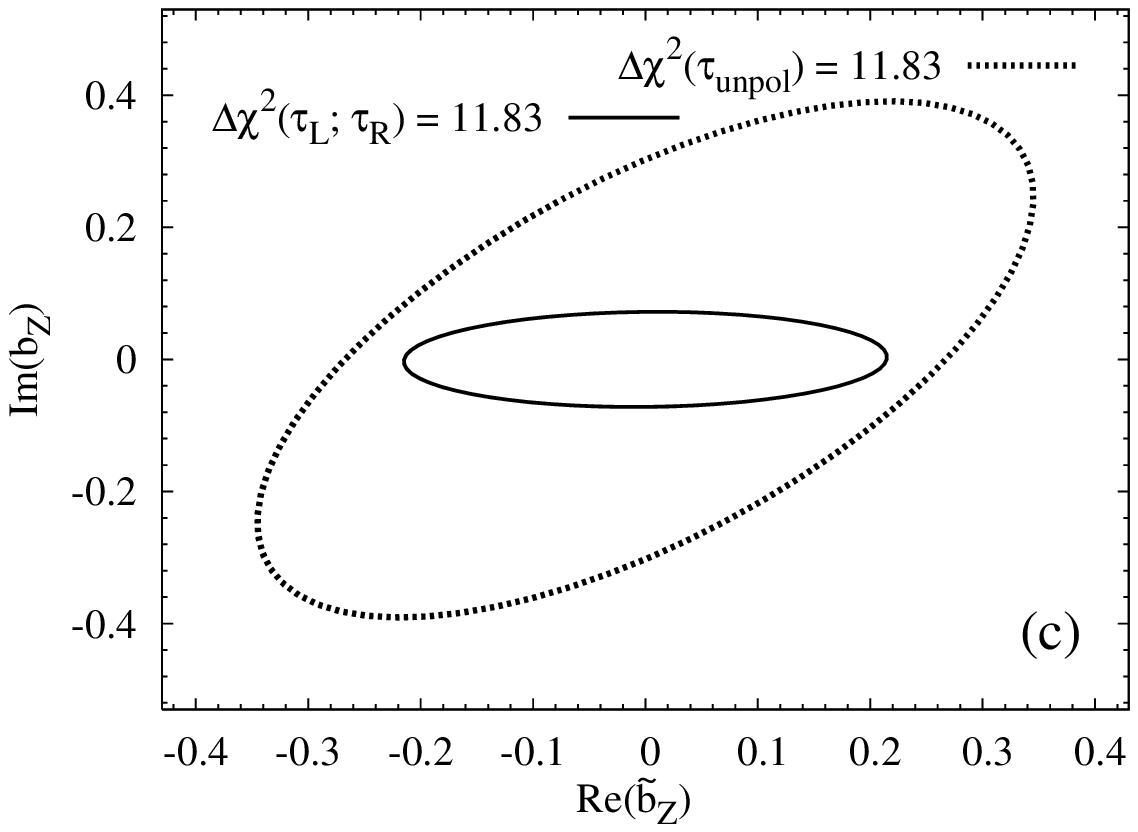,width=5.5cm,height=6.0cm}
}
\vskip -0.3cm
\caption{\em $3 \sigma$ blind regions in the 
  $\Re(\tilde b_Z)-\Im(b_Z)$ for a $\tau$-helicity isolation 
  efficiency of $40\%$. and an integrated
  luminosity of 500 fb$^{-1}$.
   Panel {\bf a} ({\bf b}) is for left-(right-)handed 
  $\tau^-$'s. 
   The horizontal, vertical and oblique lines correspond 
  to $A_{\rm{comb}}$, $A_{\rm{UD}}$ and $A{^{\prime}}_{\rm{comb}}$ 
  respectively. The ellipse combines all three and represents 
  the blind region in the plane. The inner ellipse of panel {\bf (c)} 
  combines both sets, and the outer represents the
  constraints without using $\tau$-helicity information.}
\label{fig:T-odd}
\end{figure}

Fig.~\ref{fig:T-odd} displays the region in $\Re(\tilde b_Z)-\Im(b_Z)$
plane that can be probed using the asymmetries of Eq.~\ref{asymm_tau}
for $\tau^-$'s in a specific helicity state, for an isolation
efficiency of $40\%$. Once again, a $\chi^2$ test can be 
constructed trivially. Note that  the constraint 
from $A{^{\prime}}_{\rm{comb}}$ are in opposite sense for 
left- and right-handed $\tau$'s, resulting in a rotation 
of the two $3 \sigma$ ellipses with respect to each other. 
One may then combine the information obtained from
the two final states and the results thereof are
shown in the last panel. The fact of the two individual ellipses 
being only slightly rotated with respect to the $\Re (\tilde b_Z)$ 
axes means that this exercise of combining the two leads to only 
a moderate improvement. It is interesting, though, to compare 
the result with the corresponding region when the 
$\tau$-helicity  information is unavailable. The difference 
is obvious.

\subsection{\label{sec:beamtau} Use of final state $\tau$ polarisation for 
polarised beams}
Having established that each of beam polarization and 
measurement of $\tau$-helicity can lead to substantial 
improvements, the natural question relates to the 
combination of the two effects.
Recall that the up-down asymmetry $A_{\rm
  UD}$ is proportional to $(l_e^2 -r_e^2)\,(l_\tau^2 -r_\tau^2)$ while
the combined asymmetry $A_{\rm comb}$ is proportional to $(l_e^2
+r_e^2)\,(l_\tau^2 -r_\tau^2)$. Thus, while the isolation of events with
a final state $\tau$ in a definite helicity state would 
enhance both, the use of polarised beams will only enhance $A_{\rm UD}$.
Hence in this section, we concentrate on the
improvement in sensitivity to $\Re(\tilde b_Z)$ (which is probed
by $A_{\rm UD}$).

The up-down(UD) asymmetry for $\tau$'s in negative helicity state,
with negatively polarized $e^-$ and positively polarized $e^+$ beams
(i.e. the polarisation state `$a$' as mentioned in
Sec.~\ref{sec:beamzzh}) is given by
\begin{equation}
\begin{array}{rcl}
A^{-,+}_{UD}(R1; \tau_L) &=&
\displaystyle
\frac{- 5.66 \ \Re(\tilde b_Z)}{0.836}.
\end{array}
\label{beampol_tau_L}
\end{equation}
Once again, assuming that $\tau_L$'s can be isolated 
with an efficiency of 40\% (20\%), the associated 
3 $\sigma$ limits on $\Re(\tilde b_Z)$ from this 
single measurement alone for an integrated luminosity $L= 200 \fb^{-1}$
(i.e. using only part of the data available for option (ii) of the 
luminosity divide) reads
\beq
|\Re(\tilde b_Z)| \leq  \left\{
  \barr{rcl}
0.054 & \qquad & {\rm for\,\, 40\%\,\,efficiency} 
\\[0.5ex]
0.077 && {\rm for\,\, 20\%\,\,efficiency}.
\earr
\right.
\label{eq:taupol+beampol}
\eeq
Comparing this to  Table~\ref{tab:beamlimit}, the
large gain in sensitivity is obvious. And that too with only part 
of the data. This is significantly better than 
the maximal sensitivity available for 
unpolarized beams, namely $|\Re(\tilde b_Z)| \leq 0.067$ obtainable for 
the $e^+e^-H$ final state with $R2$-cut and isolating $\tau$-helicities 
with a $40\%$ efficiency. 
Once the data for the other combinations of beam polarizations 
and the $\tau$-helicity, viz.
\begin{equation}
\begin{array}{rcl}
A^{+,-}_{UD}(R1; \tau_L) &=& \dis
\frac{4.1 \ \Re(\tilde b_Z)}{0.627}.
\\[2ex]
A^{-,+}_{UD}(R1; \tau_R) &=&
\displaystyle
\frac{4.17 \ \Re(\tilde b_Z)}{0.617}.
\\[2ex]
A^{+,-}_{UD}(R1; \tau_R) &=&
\displaystyle
\frac{- 3.02 \ \Re(\tilde b_Z)}{0.462}.
\end{array}
\label{beampol_tau}
\end{equation}
are taken into account (using a $\chi^2$ test), one obtains, for a 
40\% isolation efficiency,  
\beq
|\Re(\tilde{b}_{Z})| \leq \left\{
\barr{rcl}
0.032 & \quad & {\rm for \,\,option \,(i) }
     \\[0.5ex]
0.040 & \quad & {\rm for \,\,option \,(ii) } \ ,
  \earr
  \right.
\label{chisqlim-beampol-taupol40}
\eeq
with the numbers for an isolation efficiency of 20\% being a little 
worse. The combined use of beam polarisation and $\tau$-helicity 
measurement thus plays a very productive role, 

%

To summarize the entire section, we have demonstrated that the CP-odd
and $\tilde T$-odd $ZZH$ couplings can be probed far better by the use
of polarised beams and/or the information of final state polarisation.
The sensitivity to CP-even and $\tilde T$-even anomalous $ZZH$
couplings, ($\Delta a_Z$ and $\Re(b_Z)$), on the other hand, show only
a marginal improvement in their sensitivity limits. However, use of
polarised beams helps to constrain these couplings independent of each
other.

\section{\label{sec:wwh} Beam Polarization and the $WWH$ Couplings} %
A study of the anomalous $WWH$ couplings is possible via the process
$e^+ e^- \rightarrow \nu \bar \nu H$. However, since it receives
contributions from both the $t$-channel $WW$ fusion diagram and the
$s$-channel Bjorken diagram, this determination needs 
knowledge of the anomalous $ZZH$ couplings, which fortunately, can be
measured well from measurements of other final states.  It may be
argued that, for completely polarized $e^\pm$ beams, the cross section
$\sigma_{LR}$ gets contribution from both the Bjorken and fusion
diagrams, whereas only the first contributes to $\sigma_{RL}$, and
hence it should be possible to use cross sections with different
polarisation combinations to reduce the contamination due to the
anomalous $ZZH$ couplings. It should be noted however, that in
realistic situations one would not have $100 \%$ beam polarisation, and
thus this effect cannot be entirely neutralized.

Apart from this possible contamination, the determination of $WWH$
anomalous coupling suffers from one more limitation.  The presence of
a pair of neutrinos deprives us of the full knowledge of the momenta
of the final state fermions and, thus, does not allow construction of
$\tilde T$-odd observables.  Total rates and forward-backward
asymmetry with respect to polar angle of the Higgs boson (each for
different combinations of beam polarisations) are the only observables
available in the present case. Using the
notation of Eq.~\ref{def-pol_states} for the polarisation states, the
states `$c$' and `$d$' do not contribute to the process $e^+\, e^- \to
\nu \bar \nu H$, for the case of $100 \%$ polarisation. Even with the
assumed values of $80 \%$ and $60 \%$ polarisation for the  $e^-$ and $e^+$ 
beams respectively, these two combinations will correspond to rather small 
cross sections. Therefore, we restrict ourselves 
to the data accrued from the other two polarisation combinations 
viz. $a$ and $b$. 
On imposition of the 
$R1^\prime$- or the $R2^\prime$-cut, the cross sections are now given 
by\footnote{
As stated earlier in Sec.~\ref{sec:vvh}, we may choose $a_Z$ to be real,
without any loss of generality. But once we make this choice, 
all the $WWH$ couplings (including $\Delta a_W$) may be complex.}  

\beq
\barr{rcl}
\sigma^{-,+}(R1^{\prime};\nu) &=& 
\left[9.09 + 17.6 \ \Delta a_Z + 0.60 \ \Re(\Delta a_W) 
+ 83.8 \ \Re(b_Z) - 3.62 \ \Im(b_Z)\right. 
\\[1ex]
&& \left. -\  0.48 \ \Im(\Delta a_W)
+ \  0.90 \ \Re(b_W) + 1.51 \ \Im(b_W) \right] { \rm fb}
\\[1ex]
\sigma^{+,-}(R1^{\prime};\nu) &=& 
\left[6.63 + 13.2 \ \Delta a_Z + 0.02 \ \Re(\Delta a_W) 
+ 63.4 \ \Re(b_Z) - 0.10 \ \Im(b_Z)\right.
\\
&& \left. -\  0.01 \ \Im(\Delta a_W)
+ \  0.03 \ \Re(b_W) + 0.04 \ \Im(b_W) \right]  { \rm fb}
\\[1ex]
\sigma^{-,+}(R2^{\prime};\nu) &=& \left[
102 - 0.45 \ \Delta a_Z - 17.6 \ \Re(b_Z) - 0.31 \ \Im(b_Z)
\right. \\ 
&&
\left. +\ 205 \ \Re(\Delta a_W) - \ 38.3 \ \Re(b_W) \right]  { \rm fb}
\\[1ex]
\sigma^{+,-}(R2^{\prime};\nu) &=& \left[
3.23 + 0.78 \ \Delta a_Z + 4.31 \ \Re(b_Z)  +\ 5.68 \ \Re(\Delta a_W) - \ 1.06 \ \Re(b_W) \right]  { \rm fb}
\earr
\label{sig:nn}
\eeq

\begin{figure}[!h]
\begin{center}
\epsfig{file=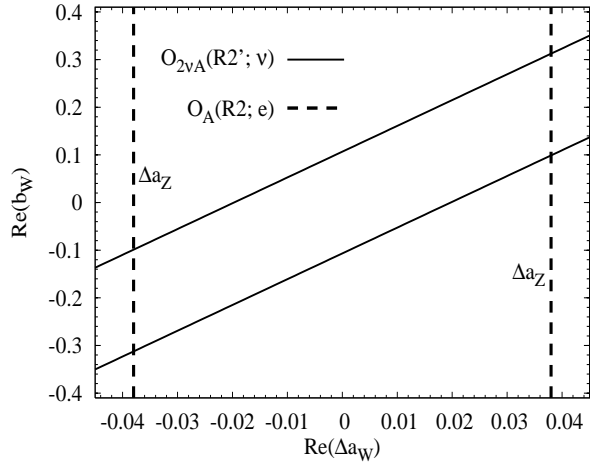,width=8.5cm,height=6.5cm}
\caption{\label{fig:aw-rebw}{\em The region in the 
$\Re(\Delta a_W)-\Re(b_W)$ plane
consistent with $3\sigma$ variations in $\mathcal O_{2\,\nu_ A}$
for an integrated luminosity of 125 fb$^{-1}$. The vertical line
shows the sensitivity limit on $\Delta a_Z$ from Fig.~\ref{fig:az-rbz}}}
\end{center}
\end{figure}

That the WWH couplings make a small contribution for 
the $(+,-)$ case is easy to understand. Furthermore, 
the contributions from $ \Im(b_{W, Z})$ are small as these
have to be proportional
to the absorptive part of the $Z$-propagator in Bjorken diagram. 
As can be expected, the sensitivity to the WWH couplings 
is enhanced if one can successfully eliminate the contribution 
of the $Z$-diagram (this has the further advantage of eliminating 
the $\nu_\mu$ and $\nu_\tau$ events). However, since even the 
$R2^{\prime}$-cut~(see Eq.~\ref{cuts:EH}) cannot eliminate the $Z$ diagram 
entirely, as is seen from Eq.~\ref{sig:nn}, we consider a 
particular linear combination of the cross sections, namely 
\beq
\barr{rcl}
\mathcal O_{2\,\nu_ A} &\equiv& 
\sigma^{-,+}(R2^{\prime};\nu) 
+ 4\ \sigma^{+,-}(R2^{\prime};\nu) 
\\
 &=& \left[ \eta_1 + 115 + 228\ \Re(\Delta a_W) - 42.5\ \Re(b_W) \right] 
    \, {\rm fb{}},
\\[1ex]
\eta_1 & \equiv & 2.66 \ \Delta a_Z - 0.36 \ \Re(b_Z) - 0.31 \ \Im(b_Z).
\earr
  \label{o2nuA} 
\eeq 
Thus the contamination from ZZH couplings is
contained entirely in $\eta_1$.
 Using the analysis of the previous
section (which did not involve the $\nu \bar\nu H$ final state), we
have $|\eta_1| \leq 0.13(0.14)$ respectively for options (i) and (ii)
of the luminosity division (Eqs.~\ref{lum_div}). In other words, the
uncertainty due to a lack of precise knowledge of the ZZH couplings is
reduced to negligible proportions enabling us to constrain a
combination of $\Re(\Delta a_W)$ and $\Re(b_W)$ virtually independent
of ZZH couplings: 
\beq |2\ \Re(\Delta a_W) - 0.37\ \Re(b_W)| \leq
0.040 \, (0.036) 
\label{simlim:aw-bw}
 \eeq 
for the two choices of luminosity division among  different 
polarisation modes concerned (see Fig.~\ref{fig:aw-rebw}). 
The relatively
small difference between the sensitivities indicates that the
luminosity division is not very crucial as long as a sufficiently
large fraction is devolved into the canonical choices of ($+, -$) and
($-, +$).  It may be also noted here that due to large value of SM 
cross section, the fluctuation in the cross section with $R2^{\prime}$-cut
given by Eq.~\ref{fluctuation} is dominated by the systematic error 
as mentioned in Sec.~\ref{sec:cuts}. 
Thus small changes in luminosity or cross section is not going to make 
much difference to the sensitivity limits of the couplings that are 
probed by cross section $\sigma^{-,+}(R2^{\prime};\,\nu)$ or 
$\mathcal O_{2\,\nu_ A}$.

While it is true that we can only probe a combination of
these two couplings using $\mathcal O_{2\,\nu_ A}$, use of beam
polarization clearly helps to reduce contamination from $ZZH$ couplings
to this determination. 
It is to be noted that since only a particular linear
combination of $\Re(\Delta a_W)$ and $\Re(b_W)$ can be
probed, limits on them are strongly correlated. 
Of course, it is possible to derive a
constraint on the orthogonal combination from the data of
Eq.~\ref{sig:nn}, but it is too weak to be of any relevance. If we
make a further assumption of only one of these couplings being
non-zero, we would obtain 
 \beq 
\barr{rclcl} |\Re(\Delta a_W)| &\leq
& 0.020 \, (0.018) & \qquad & \mbox{for } \Re(b_W) = 0 \\ 
|\Re(b_W)| &\leq & 0.110 \, (0.095) & & \mbox{for } \Re(\Delta a_W) = 0
\label{indlim:aw-bw} 
\earr 
\eeq
with the two limits corresponding to
options (i) and (ii) of Eqs.~\ref{lum_div}. Although a comparison with
the results obtained earlier~\cite{Biswal:2005fh} shows only a
marginal improvement in the individual limits, the determination is
now free of uncertainty coming from contamination from the $ZZH$
couplings.

At this stage, it is intriguing to consider 
the consequences of an additional assumption 
(made in Ref.\cite{Biswal:2005fh})
of $\Delta a_W = \Delta a_Z $, which is 
found to be true in some cases~\cite{pilaftsis-wagner} and would be motivated 
by $SU(2) \times U(1)$ invariance of the 
effective theory. By itself, this would 
impose a bound on $\Delta a_W$, courtesy Eq.~\ref{lim_az} and 
hence that would lead to a closed area as shown in Fig.~\ref{fig:aw-rebw}. 
However, one should also note 
that $SU(2) \times U(1)$ invariance would further equate $b_W$ and $b_Z$, and 
that the correlation between constraints in the 
$\Re(b_W)$--$\Re(\Delta a_W)$ plane 
(Fig.~\ref{fig:aw-rebw}) is in the opposite sense to that in the 
$\Re(b_Z)$--$\Delta a_Z$ plane (Fig.~\ref{fig:az-rbz}). Thus, an assumption 
of such an invariance would lead to far stronger constraints.
In particular, the improvement is dramatic for $\Re(b_W)$.

As for the other CP-even $WWH$ couplings, namely 
$\Im(\Delta a_W)$ and $\Im(b_W)$, note that 
their contributions arise 
from the interference of the $WW$-fusion diagram with the absorptive part of
the $Z$-propagator in the Bjorken diagram.  Hence the
corresponding terms appear only in total rate with $R1^\prime$-cut and
being proportional to the width of $Z$-boson, are very small. Since 
the presence of two neutrinos in the final state does not allow us to
construct any $\tilde T$-odd observables, this study is virtually
insensitive to these two couplings. Of course, an assumption of 
$SU(2) \times U(1)$ invariance would change matters drastically.

What remains is to investigate $\tilde b_W$, and this being 
a CP-odd coupling, can be probed through the
forward-backward (FB) asymmetry with respect 
to the polar angle of Higgs boson. These asymmetries, 
$R1^\prime$- or and $R2^\prime$-cuts can be expressed as 
\beq
\barr{rcl}
A_{FB}^{-,+}(R1^{\prime};\nu) 
&=& \dis \frac{1}{9.09}\left[- 2.29 \ \Re(\tilde b_Z) - 36.9 \ 
  \Im(\tilde b_Z)
+ 0.57 \ \Re(\tilde b_W) - 0.47 \ \Im(\tilde b_W)\right],
\\[2ex]
A_{FB}^{+,-}(R1^{\prime};\nu) &=& \dis 
\frac{1}{6.63}\left[- 0.064 \ \Re(\tilde b_Z) + 27.1 \ \Im(\tilde b_Z)
+ 0.02 \ \Re(\tilde b_W) - 0.01 \ \Im(\tilde b_W)\right],
\\[2ex]
A_{FB}^{-,+}(R2^{\prime};\nu) &=& \dis
\frac{1}{102}\left[5.17 \ \Im(\tilde b_Z) + 7.83 \ \Im(\tilde b_W) \right],
\\[2ex]
A_{FB}^{+,-}(R2^{\prime};\nu) &=& \dis \frac{1}{3.23}\left[2.1 \ \Im(\tilde
b_Z) + 0.22 \ \Im(\tilde b_W) \right].
\earr
\label{asm:nnh}
\eeq
where the numbers once again represent the corresponding 
cross sections in femtobarns. With the last two of 
Eqs.~\ref{asm:nnh} involving just a single 
$WWH$ coupling, these can be combined to eliminate 
the remaining dependence on the $ZZH$ anomalous couplings to leave us 
\beq
\barr{rcl}
A_{2_{\rm mix}} & \equiv &  13 \ A_{FB}^{-,+}(R2^{\prime};\nu) - 
A_{FB}^{+,-}(R2^{\prime};\nu) \\
   &=& 0.009 \ \Im(\tilde b_Z) + 0.93 \ \Im(\tilde b_W).
\earr
\label{A2mix}
\eeq
The contribution  to $A_{2_{\rm mix}}$ from  $\Im(\tilde b_Z)$ 
(which, incidentally, can be probed very accurately,
 see Table~\ref{tab:beamlimit}) may be neglected, leading to 
\beq
|\Im(\tilde b_W)| \leq 0.50 \, (0.44)
\label{lim:imtilbw}
\eeq
for ${\cal L}$ = 125 \, (200) fb$^{-1}$
for each of the $(+, -)$ and $(-, +)$ polarization combinations. 
While it may seem that the improvement is marginal when compared
to the sensitivity achievable with unpolarised beams---
$|\Im(\tilde b_W)| \leq 0.46 $ for a total luminosity of
$500$ fb$^{-1}$\cite{Biswal:2005fh}---note that, 
unlike the older analysis, the current sensitivity is 
independent of any other coupling. 
In other words, the use of beam polarisation has allowed 
construction of an observable that can isolate the contribution of 
$\Im(\tilde b_W)$.

The anomalous coupling $\Re(\tilde b_W)$ being a $\tilde T$-odd
coupling has no contribution to FB asymmetries with $R2^\prime$-cut
and only a small contribution to FB asymmetries with $R1^\prime$-cut
through  the interference of the $WW$-fusion diagram with the
absorptive part of Bjorken diagram. Hence these asymmetries are not a
good probe of $\Re(\tilde b_W)$. Thus the process under consideration
can not be used to probe any of the $\tilde T$-odd couplings in the $WWH$ 
vertex. 

\begin{table}[!h]
\begin{center}
\begin{tabular}{||cc||c|c|c||c|c||}
\hline\hline
\multicolumn{5}{||c||}{Using Polarized Beams} &
\multicolumn{2}{|c||}{Unpolarized States} \\[2mm]
\hline
Coupling & &
\multicolumn{2}{|c|}{Limits (for given $\cal L$)}
&
$\begin{array}{c}
\mbox{Observables}~ \cr
\mbox{used}
\end{array}$
& Limits &
$\begin{array}{c}
\mbox{Observables}~ \cr
\mbox{used}
\end{array}$
\\
\cline{3-4}
&&
200 fb$^{-1}$&
125 fb$^{-1}$
&
&& \\
\hline
\hline
&&&&&&\\
$|\Re(\Delta a_{W})|$
& $\leq$ &0.018& 0.020 
&
$\begin{array}{l}
\mathcal O_{2\,\nu_ A}\cr
\end{array}$
& 0.019 &
$\begin{array}{l}
\sigma^{unpol}(R2^{\prime};\nu) \cr
\end{array}$\\[2mm]
\hline
\hline
&&&&&&\\
$|\Re(b_{W})|$
& $\leq$ &0.095& 0.11 
&
$\begin{array}{l}
\mathcal O_{2\,\nu_ A}\cr
\end{array}$
& 0.10 &
$\begin{array}{l}
\sigma^{unpol}(R2^{\prime};\nu) \cr
\end{array}$\\[2mm]
\hline
\hline
&&&&&&\\
$|\Im(\tilde b_{W})|$ & $\leq$ &0.44& 0.50 & 
$\begin{array}{l}
A_{2_{\rm mix}} \cr
\end{array}$
& 0.40 &
$\begin{array}{l}
A_{FB}^{unpol}(R2^{\prime};\nu) \cr
\end{array}$\\[2mm]
\hline
\hline
\end{tabular}
\vskip -0.2cm
\caption{\label{tab:simlu_lim-wwh}{\em $3\, \sigma$ limits on
anomalous $WWH$ couplings from various observables with polarized and
unpolarized beams.}}
\end{center}
\end{table}

To summarize the results of this section:  use of beam polarisation 
allows us to obtain limits on $\tilde T$-even couplings $\Re(\Delta
a_W)$, $\Re(b_W)$ and $\Im(\tilde b_W)$ {\it independent} of the $ZZH$
couplings, these being listed in 
Table \ref{tab:simlu_lim-wwh}.
 Using  linear combinations of our observables corresponding 
to different polarisation combinations for initial states, the
contamination from $ZZH$ coupling can be  reduced to negligible amount,
something that was not possible with the unpolarised beams. Further,
even though this measurement uses only two of the combinations, an equal
division of the total luminosity among all four already gives optimal
results. The limits
on $\Re(\Delta a_W)$ and $\Re(b_W)$ are highly correlated whereas
$\Im(\tilde b_W)$ is constrained independent of any other
coupling. There is no observable to constrain the $\tilde T$-odd
couplings and thus we conclude that this process is not a good probe
for these couplings. Hence one has to look to other processes to probe
these couplings. For example, this problem
may be overcome in the $e \, \gamma$ collider as shown in 
Ref.~\cite{egamma}. The WWH couplings are not contaminated by the ZZH
couplings in the process studied there viz. ($e^- \gamma \to \nu\, W^- \,
H$). Using this process in conjunction with the $P \tilde T$
conjugate process, the authors were able to construct observables that
depend only on one of the anomalous WWH couplings and hence the limits
obtained on each of the WWH couplings were independent of the others. The
anomalous VVH couplings at $e \, \gamma$ collider have been also
studied in Ref.~\cite{Sahin:2008jc}. Finally, an assumption of $SU(2) \times 
U(1)$ invariance would drastically improve the constraints for virtually 
all the couplings, whether they be $WWH$ or $ZZH$.

\vspace{0.5 cm}

\section{\label{sec:sensitivity} Dependence on $\sqrt{s}$ and 
inclusion of ISR/Beamstrahlung Effects}

All our analysis so far has been performed for a fixed centre of mass
energy, namely $\sqrt{s}= 500$~GeV. Clearly, the total cross section is
a function of energy, and the functional form may depend on the
presence of (and the identity of) any anomalous coupling, owing to
their higher-dimensional nature.  Thus, the sensitivities could, in
principle, depend on the choice of $\sqrt{s}$. Furthermore, for the
processes involving both $s$- and $t$-channel diagrams, the relative
importance of these two parts of the amplitude is, in fact, energy
dependent, and the generic enhancement of the $t$--channel
cross section with increasing beam energies may, in principle, lead to
an improvement in the sensitivity to those couplings primarily
constrained using observables with $R2$-$(R2^{\prime})$-cut. 

Even for a nominally fixed beam energy, the $\sqrt{s}$ available 
to an individual hard scattering event is generically less than this 
value owing to the ubiquitous initial state radiation (ISR)---
which is nothing but the bremsstrahlung  
radiation by the incoming particles---or 
beamstrahlung, which is a name for the 
radiation from the beam particles 
due to its interaction with the (strong) electromagnetic fields 
caused by the dense bunches of the opposite charge in a collider 
environment. Consequently, it is important to investigate the 
possibly detrimental effects of such eventualities. 

\begin{figure}[!h] 
\begin{center}
\epsfig{file=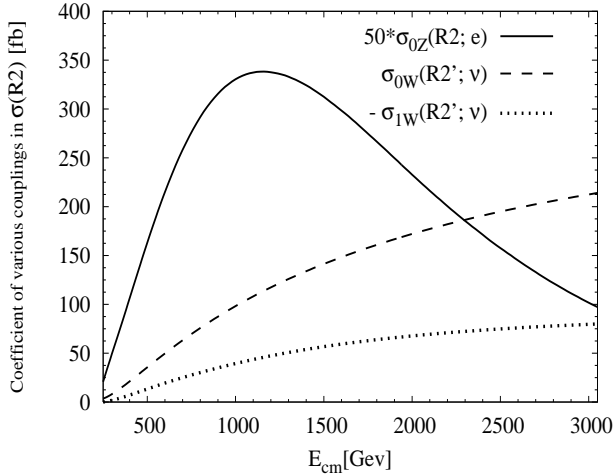,width=8.5cm,height=6.5cm}
\caption{\label{rtsvar_signnh} {\em
$\sqrt{s}$-variation of ($\sigma_{0Z}(R2;\,e)$ (the SM part in the 
cross section $\sigma(R2;\,e) = \sigma ( e^+ e^- \to e^+ e^- H) $ 
with $R2$-cut), $\sigma_{0W}(R2^{\prime};\,\nu)$ and 
$\sigma_{1W}(R2^{\prime};\,\nu)$ 
(the coefficients of  $\Re(\Delta a_W)$ and $\Re(b_W)$ respectively 
in the cross section for 
$ e^+ e^- \to \nu \bar\nu H$ with $R2^{\prime}$-cut). }} 
\end{center}
\end{figure}

To this end, we begin by studying the $\sqrt{s}$-dependence of the
observables used in this paper. For simplicity, we shall restrict 
ourselves, in this section, to unpolarised 
scattering, with the results for polarised beams expected to be similar.
Moreover, we shall concentrate on observables defined
with the $R2$-$(R2^{\prime})$-cut. For example, the $ZZH$ coupling
$\Delta a_Z$ is best probed by the total cross section with $R2$-cut for
electron final state i.e. $\sigma(R2;\,e) = \sigma ( e^+ e^- \to e^+
e^- H) $ with $R2$-cut. Fig.~\ref{rtsvar_signnh} shows the variation
of the SM part of this cross section ($\sigma_{0Z}$ of
Eq.~\ref{def-cross}) with $\sqrt{s}$.\footnote{Note that $\sigma_{0Z}$ 
has been scaled by a factor of 50 to fit in the same figure.}  
Recall that $\Delta a_Z$ simply
rescales the SM part of the cross section (Eq.~\ref{def-cross}) and
hence the sensitivity is determined simply by the corresponding number
of events, namely it scales as $N_{0Z}^{-1/2}$.
With $\sigma_{0Z}(R2;\,e)$ having a maximum at 
$\sqrt{s} \approx 1.1$ TeV, this would be the optimal beam energy 
to probe $\Delta a_Z$, with the maximal improvement (compared 
to the results quoted above) being by about 40\%.

Fig.~\ref{rtsvar_signnh} also shows the $\sqrt{s}$-variation of the SM
and $\Re(b_W)$ contributions (i.e. $\sigma_{0W}$ and $\sigma_{1W}$
respectively) to the cross section $\sigma(R2^{\prime}; \nu ) = \sigma
( e^+ e^- \to \nu \bar\nu H) $ with $R2^{\prime}$-cut which is the
best probe for the $WWH$ couplings $\Re(\Delta a_W)$ and $\Re(b_W)$.
While the monotonic increase in $\sigma_{1W}$ would seem to suggest 
that increasing $\sqrt{s}$ would readily lead to an improvement 
in the sensitivity to  $\Re(b_W)$, note that this increase saturates and, 
furthermore, that $\sigma_{0W}$ increases at least as fast. Consequently,
for moderate changes in $\sqrt{s}$, any improvement or otherwise is expected 
to be marginal at best.

\begin{figure}[!h]
 \centerline{
\epsfig{file=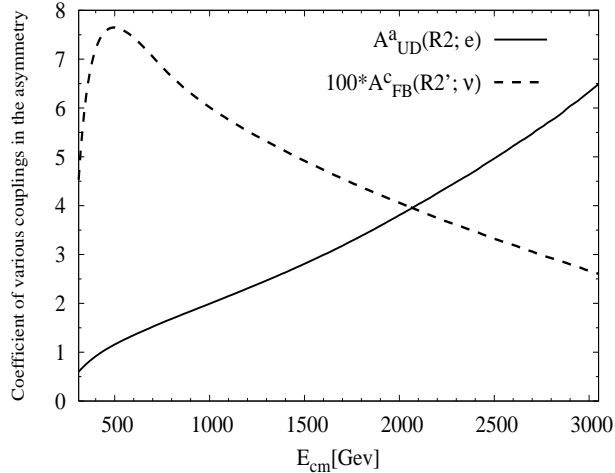,width=8.5cm,height=6.5cm}}
\caption{\label{rts_var_asymm} {\em $\sqrt{s}$-variation of the coefficients 
$A_{UD}^a$ in up-down   asymmetry and the $A_{FB}^c$ in
  forward-backward asymmetry  as defined in Eq.~\ref{def_asymm}
 of the text. Both plots are
  with $R2$-($R2^{\prime})$-cut (de-selecting Z pole) imposed. }}  
\end{figure}

Moving to asymmetries, 
the up-down asymmetry $A_{UD}(R2;\,e)$ of the final state fermion with
respect to the $H$-production plane, 
and the forward-backward asymmetry 
with respect to polar angle of the Higgs boson, 
$A_{FB}(R2^{\prime};\,\nu)$
 have been used to constrain 
$\Re(\tilde b_Z)$ and $\Im(\tilde b_W)$ respectively~\cite{Biswal:2005fh}. 
Expressing these as
\bea
A_{UD}(R2;\,e) &=& A_{UD}^a \,\Re(\tilde b_Z)  \nn\\ 
A_{FB}(R2^{\prime};\,\nu)&=& A_{FB}^b \,\Im(\tilde b_Z) +   
A_{FB}^c \,\Im(\tilde b_W) 
\label{def_asymm}
\eea 
the coefficients $A^a_{UD}$ and $A^c_{FB}$ 
are plotted in Fig.~\ref{rts_var_asymm} as a function of 
$\sqrt{s}$.  
 The figure clearly shows that
the sensitivity to
$\Re(\tilde b_Z)$ is expected to improve at higher $\sqrt{s}$, while, 
$\sqrt{s} = 500$~GeV is an optimal choice for $\Im(\tilde b_W)$ and higher 
energies would only tend to deteriorate the sensitivity.
The arguments above  are reflected by Table
\ref{tab:high-limit} which summarizes the sensitivity limits
at the $3\,\sigma$ level,  with an integrated luminosity
of 500 fb$^{-1}$, for different center of mass energies. 
%
\begin{table}[!h]
\begin{center}
\begin{tabular}{|cccccc|}
\hline
\hline
Coupling& &
$\begin{array}{l}
3~\sigma~\mbox{limit at}\cr
\sqrt{s}~=~0.5 \cr
\mbox{~~~~TeV}
\end{array}$
&
$\begin{array}{l}
3~\sigma~\mbox{limit at}\cr
\sqrt{s}~=~1 \cr
\mbox{~~~~TeV}
\end{array}$
&
$\begin{array}{l}
3~\sigma~\mbox{limit at}\cr
\sqrt{s}~=~3 \cr
\mbox{~~~~TeV}
\end{array}$
&
$\begin{array}{c}
\mbox{Observable}\cr
\mbox{used}
\end{array}$
\\ \hline
&&&&&\\[-3mm]
$|\Delta a_Z|$ & $\leq$ & 0.040 & 0.030 &0.049&
$\sigma(R2;e)$ \\[2mm]
& & 0.043 & 0.031 &0.039& $\sigma^{\mbox{\small
ISR+Beam.}}(R2;e)$ \\[2mm]
\hline
&&&&&\\[-3mm]
$|\Re(\tilde b_Z)|$ & $\leq$ & 0.067 & 0.028 &0.015&
$A_{UD}(R2;e)$ \\[2mm]
& & 0.075 & 0.032 &0.018& $A^{\mbox{\small
  ISR+Beam.}}_{UD}(R2;e)$ \\[2mm]
\hline
&&&&&\\[-3mm]
$|\Re(\Delta a_W)|$ & $\leq$ & 0.019 & 0.016
&0.016 & $\sigma(R2^{\prime};\nu)$ \\ [2mm]
& & 0.019 & 0.017 &0.016
& $\sigma^{\mbox{\small ISR+Beam.}}(R2^{\prime};\nu)$ \\ [2mm]
\hline
&&&&&\\[-3mm]
$|\Re(b_W)|$ & $\leq$ & 0.10 & 0.082
&0.084 & $\sigma(R2^{\prime};\nu)$ \\ [2mm]
& & 0.11 & 0.084 &0.083
& $\sigma^{\mbox{\small ISR+Beam.}}(R2^{\prime};\nu)$ \\ [2mm]
\hline
&&&&&\\[-3mm]
$|\Im(\tilde
b_W)|$ & $\leq$ & 0.40 & 0.42 &0.89& $A_{FB}(R2^{\prime};\nu)$ \\ [2mm]
& & 0.43 & 0.41 &0.71& $A^{\mbox{\small ISR+Beam.}}_{FB}(R2^{\prime};\nu)$
\\ [2mm] \hline \hline
\end{tabular}
\caption{\label{tab:high-limit}{\em Individual sensitivity limits (assuming only the relevant coupling to be non-zero) at the 
$3\sigma$ level for an integrated luminosity of 500 fb$^{-1}$ on 
various anomalous couplings at different c.m. energies without and 
with the inclusion of ISR and beamstrahlung effects. 
The results correspond to unpolarised beams.}}
\end{center}
\vskip -0.2cm
\end{table}
%

The above analysis did not take into account
either of ISR and beamstrahlung. 
We now proceed 
to do so, using the structure function formalism \cite{Drees:1992ws} 
to incorporate these effects. The differential scattering 
cross section for a given process
$e^-(p_1) + e^+(p_2) \to X  (\gamma)$ can be expressed as:
\[
d\sigma[e^+e^- \to X (\gamma)] 
=  f_{e/e}(x_1) \; f_{e/e}(x_2) \;
d\hat{\sigma}[e^+e^- \to X] 
(\hat{s})\ ,
\label{eq:diff_isr}
\]
where the electron luminosity function $f_{e/e}(x)$ 
describes the probability of finding an electron with a momentum fraction
$x$ of the nominal beam energy, or, in other words, the 
probability with which an electron
energy $E = \sqrt{s}/2$ emits one or more photons with total
energy $(1-x) E$ resulting in the reduction of its energy to $E_e = x
E$.  Hence the square of the effective c.m. energy $\hat{s}$ can be
expressed as: $\hat{s}$ $\simeq x_1 x_2 s$.  

Using the Weisz\"acker-Williams approximation, 
one can write down  the luminosity function for
ISR as~\cite{Kuraev:1985hb}
\begin{eqnarray}
f^{\rm ISR}_{e/e}(x) = \frac{\beta}{16}
\left[ (8 + 3\beta) (1 - x)^{\beta/2 - 1} - 4 (1 + x) \right] \ ,
\label{eq:ISR}
\end{eqnarray}
where
\begin{eqnarray}
 \beta = \frac{2 \alpha_{em}}{\pi}\left(\log \frac{s}{m_e^2} - 1 \right) \ ,
\end{eqnarray}
and $\alpha_{em}$ is the fine-structure constant.

The beamstrahlung spectrum depends on 
electron beam energy $E$, and the parameters 
such as the number of electrons per bunch $N_e$, 
the bunch dimensions (for a Gaussian bunch profile) in both 
the longitudinal direction ($\sigma_z$) as well as 
in the transverse directions ($\sigma_{y, x}$). It is convenient to 
introduce a ``beamstrahlung parameter'' $\Upsilon$~\cite{Chen:1991wd}
given by:
$$\Upsilon = \frac { 5 \, r_e^2 \, E \, N_e } 
                   { 6 \, \alpha_{em} \, \sigma_z \, 
                             \left(\sigma_x + \sigma_y \right) \, m_e } \ , 
$$
where $r_e$ is the classical electron radius, and $m_e$ its mass.
The electron spectrum $f^{\rm beam}_{e/e}(x)$ describing the effects of 
beamstrahlung,  can be written in a closed analytical form for 
$\Upsilon \lsim\ 10$ ~\cite{Chen:1991wd}. 
Armed with all these, the expression for the electron spectrum function, 
including both ISR and beamstrahlung effects, may be written 
as~\cite{Drees:1992ws}:
\begin{eqnarray}
f_{e|e}(x) = \int_x^1 ~\frac{d\xi}{\xi} ~f_{e|e}^{\rm ISR}(\xi)
~f_{e|e}^{\rm beam}
\left(\xi^{-1} \, x\right),
\end{eqnarray}
where $f_{e|e}^{\rm ISR}(\xi)$ is as given in Eq.~\ref{eq:ISR} above, 
whereas $f_{e|e}^{\rm beam}\left(\xi^{-1} \, x\right)$ is that given 
by Eq. 22 of Ref.~\cite{Chen:1991wd}.

We now analyse how the ISR and beamstrahlung effects modify the
sensitivity limits of $\Re(\tilde b_Z)$, for which we had observed
an improvement in sensitivity at higher $\sqrt{s}$ as listed in 
Table~\ref{tab:high-limit}. 
In our analysis we use the beamstrahlung parameters to
be~\cite{Assmann:2000hg} 
\bea \sigma_z &=& 30 \, {\rm \mu m\quad for\quad all \quad energies}, \nn
\\ 
\Upsilon &=& 0.3,\,\, 1.0 \,\,\, {\rm and} \,\,\, 8.1 \quad {\rm
  for }\quad E_{\rm cm} = 0.5,\,\, 1.0,\,\,\, {\rm and}\,\,\, 3.0\,
   {\rm TeV} \quad {\rm respectively}.
\label{beam_para}
 \eea

\begin{figure}[!h]
\centerline{
\epsfxsize=8.0cm \epsfysize=7.5cm \epsfbox{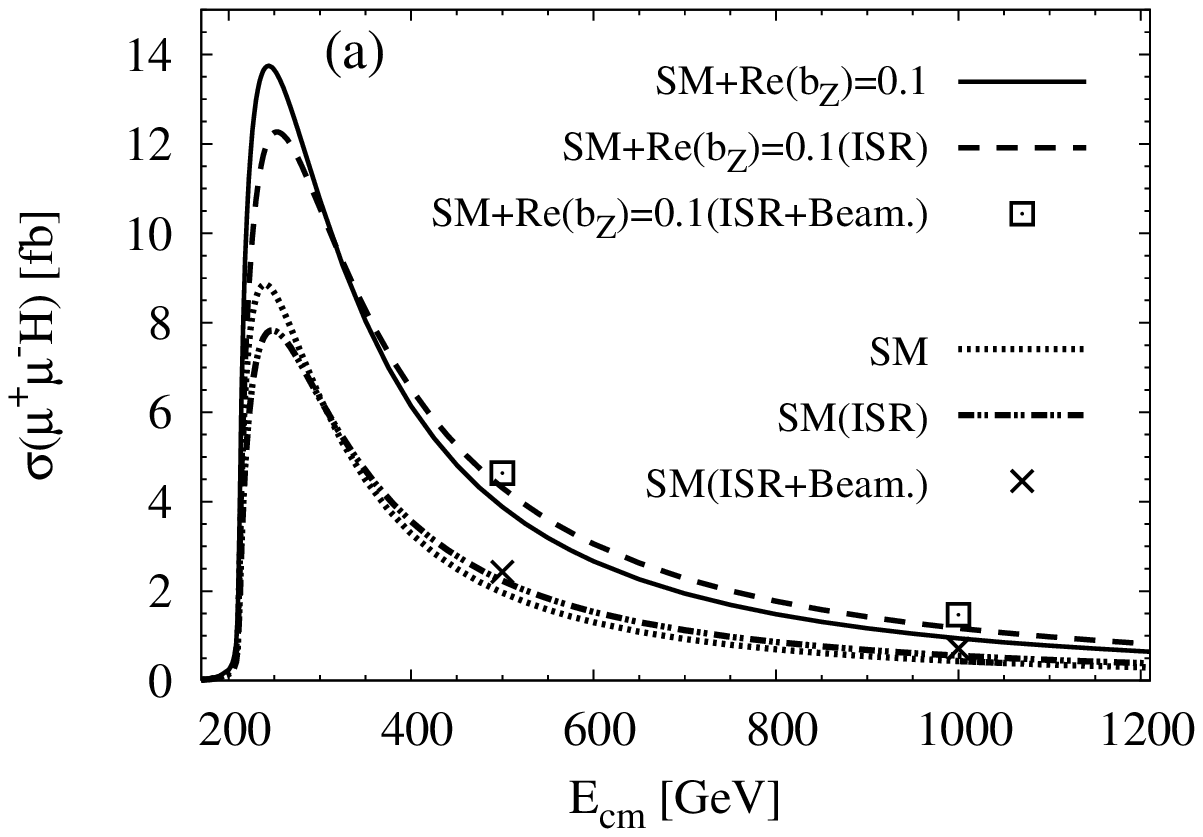}
\epsfxsize=8.0cm \epsfysize=7.5cm \epsfbox{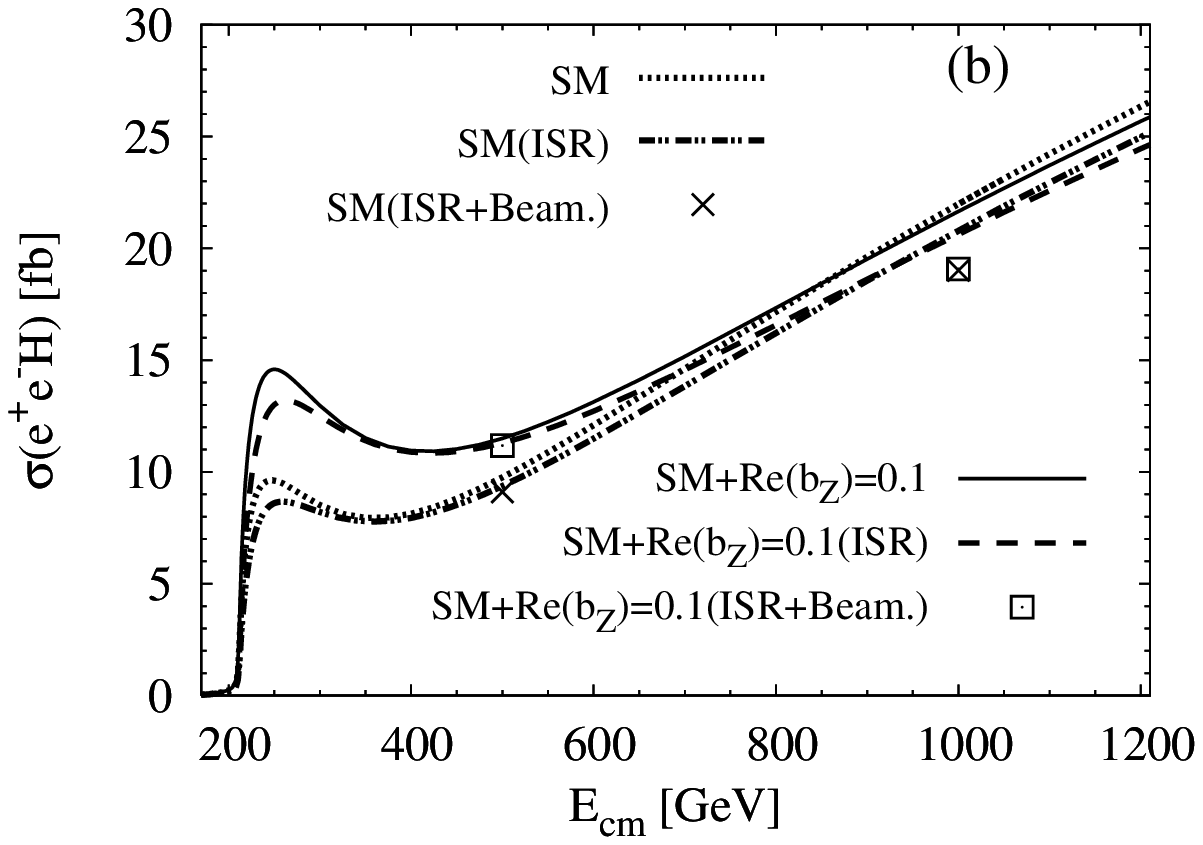}}
\caption{\label{isr_rtsvar}{\em $\sqrt{s}$-variation of cross sections
    with and without the ISR effects
    for final states with muons or electrons.
    The figure also shows the points at $\sqrt{s}=
    0.5$~TeV and 1.0~TeV when beamstrahlung effects are also included.
    Beamstrahlung effects are negligible at 
    $\sqrt{s}= 0.5$~TeV. }}
\end{figure}

Fig.~\ref{isr_rtsvar} compares the total rates with and without ISR
effect for different final states. 
The effective c.m. energy $\hat{s}$ of the electron beam decreases due
to ISR.  As a result, away from the threshold, the ISR effects
increase the rates of pure $s$-channel processes, such as $e^+\, e^-
\to \mu^+\,\mu^-\, H$. Near the threshold, however, the cross section
rises with the effective $\sqrt{s}$ and hence the ISR effects cause a
decrease in the rate.  For the process under discussion, the
cross-over from one behaviour to the other takes place at about $300$
GeV.  On the other hand, ($e^+ e^- H$)-production receives both $s$-
and $t$- channel contributions, and hence there is no such cross-over
with ISR effects always decreasing the rates.  We thus expect these effects
to reduce somewhat the improvement observed above in probing
$\Re(\tilde b_Z)$ at higher energies.

At $\sqrt{s}$ = 500~GeV, beamstrahlung effects are negligible whereas
ISR effects cause 10\% to 15\% change in cross section 
(see Fig.~\ref{isr_rtsvar}). 
However, this does not cause significant changes in sensitivities 
as  both the SM and anomalous
parts are affected  in much the same way.  At higher c.m. energies 
both ISR and beamstrahlung effects are nontrivial and significant.  
However, once again, the effect is in the same direction, and roughly 
of the same magnitude, for the SM and in the presence of anomalous couplings
(Fig.~\ref{isr_rtsvar}). Although the figure demonstrates this for 
total cross sections, the story is similar even for partial cross sections
and for the other couplings as well.

Since the constraints on 
$\Re(\tilde b_Z)$ are the only ones to improve significantly 
with increasing $\sqrt{s}$ (see Table \ref{tab:high-limit}), 
we concentrate on the effects of ISR and beamstrahlung on this coupling. 
With $\Re(\tilde b_Z)$ being best probed by 
the up-down asymmetry (with $R2$-cut) for electrons in ($e^+ e^- H$) 
production, the ISR and beamstrahlung effects can be summarised 
as listed in Table~\ref{tab:UD_isr_beam}.
\begin{table}[!h]
\begin{center}
\begin{tabular}{||c||c|c|c||}
\hline\hline
$\sqrt{s}$&\multicolumn{3}{c||}{UD-asymmetry($A_{UD}(R2;\,e)$)} \\[2mm]
\cline{2-4} &No ISR \& Beam.&With ISR & With ISR \& Beam.\\
\hline
\hline
&&&\\%
0.5~TeV &1.16 $\Re(\tilde b_Z)$ & 1.13 $\Re(\tilde b_Z)$
& 1.1 $\Re(\tilde b_Z)$ \\
&&&\\%
\hline
\hline
&&&\\%
1~TeV &2.00 $\Re(\tilde b_Z)$ & 1.94 $\Re(\tilde b_Z)$
& 1.85 $\Re(\tilde b_Z)$ \\
&&&\\%
\hline
\hline
&&&\\%
3~TeV &6.29 $\Re(\tilde b_Z)$ & 5.60 $\Re(\tilde b_Z)$
& 4.23 $\Re(\tilde b_Z)$ \\
&&&\\%
\hline
\hline
\end{tabular}
\vskip -0.2cm
\caption{\label{tab:UD_isr_beam} {\em
Up-down asymmetry for final state electron for
$R2$-cut($A_{UD}(R2;\,e)$)
with and without ISR and beamstrahlung effects
at different $\sqrt{s}$'s.}}
\end{center}
\end{table}

As is readily seen, the effects are almost negligible even for 
$\sqrt{s} = 1$ TeV, and only marginally important for 
$\sqrt{s} = 3$ TeV. The consequent shift in the sensitivities 
are summarised in Table \ref{tab:high-limit}. 
The smallness of the effects can be understood by realizing that, 
on the imposition of the $R2$-cut (de-selecting the $Z$-pole), 
the energy dependence of the cross sections (total or partial) is only 
logarithmic. Further, both SM and anomalous parts have a similar 
dependence. Hence 
although the cross sections are affected significantly there is little 
effect on the sensitivity limits of the anomalous parts after inclusion 
of ISR and beamstrahlung effects.

We conclude this section by making a few general observations :
\begin{itemize}
\item With increasing energy, the observables with the $R2$-cut imposed 
gain more in sensitivity as compared to those defined with the $R1$-cut.
\item  
Observables with $R1$-cut (selecting $Z$-pole) are affected more by ISR
and beamstrahlung corrections because of the usual $s$-channel suppression,  
whereas the observables with $R2$-cut
(de-selecting $Z$-pole) have only logarithmic $\sqrt{s}$-dependence 
and hence do not suffer as significant corrections.
\item By using higher c.m. energies, we gain significantly in the
  sensitivity to $\Re(\tilde b_Z)$ (upto a factor of 2 at
  $\sqrt{s} = 1$~TeV as compared to the case of 500~GeV). 
Neither the ISR nor the beamstrahlung effects change the 
sensitivity significantly.
\item
There is no significant gain in the sensitivity to anomalous WWH
couplings even at very high energies.
\item
In totality, we do not gain much in the sensitivity by going
 to higher c.m. energies. Thus, running the collider at lower 
energies (say $\sqrt{s} = 500\,$~GeV), but with polarised beams is 
more beneficial for the study of anomalous VVH couplings. 
\end{itemize}

\section{\label{sec:conclude}Summary}
With the plethora of the gauge boson couplings with a Higgs 
that an effective theory allows, it is always going to 
be a complicated task to unravel these. 
It was shown in Ref.~\cite{Biswal:2005fh} that it is possible to construct
observables with definite $CP$ and $\tilde T$ transformation
properties that are sensitive to a single anomalous coupling (the one
with matching $CP$ and $\tilde T$ properties) and hence these
observables may be used to obtain robust constraints on the couplings
in a model independent way. With polarised beams, we have the
possibility of constructing even more such observables.

We have performed our analysis by imposing kinematical cuts on
different final state particles to reduce backgrounds. In addition, we
also include the reduction (in the event rates) caused by considering
only the events wherein the $H$ decays into a $b \bar b$ pair (with
branching fraction $\sim 0.68$), with a realistic $b$-tagging
efficiency of 70\%. This renders, to a great extent, our estimates of
sensitivity quite realistic.

Almost all of the asymmetries pertaining to the probe of 
the $ZZH$ vertex, that were constructed in Ref.~\cite{Biswal:2005fh}
for unpolarised beams and neglecting the polarisation of
the final state particles, are proportional to the difference between
the squared right and left handed coupling of the $Z$ to charged
leptons.  Consequently, one expects the sensitivity to the
corresponding anomalous couplings to be enhanced once one uses a
specific polarisation, thereby overcoming this cancellation between
the (almost) equal left and right handed couplings of the $Z$ boson to
leptons. Indeed, we observe this.  
For example, the use of beam polarization alone leads 
on the one hand, to the disentangling of the couplings
$\Delta a_Z$ and $\Re(b_Z)$ (not possible for 
unpolarised beams), and on the other, to an 
improvement in the sensitivities to $CP$-odd $ZZH$ couplings
by a factor of upto 5-7. 

With the helicity of a $\tau$ in the final state decipherable from 
the momentum distribution of the charged prongs in its decay, 
one can construct asymmetries pertaining to $\tau$'s 
with a specific helicity. Assuming that it would  
be possible to 
isolate events with $\tau$'s in a definite helicity
state with an efficiency of (say) $40 \%$, we use  
such asymmetries to demonstrate that sensitivity to 
$\Im(b_Z)$ and $\Re(\tilde b_Z)$ could be improved 
by a factor of about $3$. While an earlier optimal observable
analysis~\cite{Hagiwara:2000tk} had indeed investigated 
the use of $\tau$ helicity, our analysis differs in that 
we have constructed an observable that 
can be measured in a simple counting experiment and catches the
essence of the optimal observable.  Also worth noting is 
that the use of final
state $\tau$ measurement along with polarised beams, allows us to
improve on the sensitivity for $\Re(\tilde b_Z)$ by a factor of about 2.

As far as the constraints on the anomalous
$WWH$ couplings are concerned, the problem of a smaller number of
observables due to the presence of missing neutrinos in the the $\nu
\bar \nu H ( b \bar b)$ still exists. However, by constructing new
observables that are combinations of the observables corresponding to
different polarisation states, the contamination from the anomalous
$ZZH$ couplings is highly reduced in case of $\tilde T$-even $WWH$
couplings which is a great advantage over the use of unpolarised
beams. For $\tilde T$-odd $WWH$ couplings though, we conclude that this
process is not a good probe for them because of
non-availability of any $\tilde T$-odd observable.  Hence one has to
look to other processes to probe these couplings.

We have also investigated possible gains in sensitivity by going to
higher centre of mass energies. This though renders ISR and the
beamstrahlung effects progressively more important.  As a matter of
fact, even for $\sqrt s = 500$ GeV, the ISR effects can affect
cross sections with the $R1$-cut by $10$--$15\%$.  Fortunately, the ISR and
beamstrahlung effects on sensitivity limits is very modest, as both
the standard model and the anomalous contributions change by similar
amounts causing only small numerical effects, especially on the
asymmetries.
As for  the variation in sensitivity with $\sqrt{s}$ 
ranging from 300 GeV to 3 TeV, the observables with the $R1$-cut (selecting 
the $Z$-pole) are naturally associated with a larger 
cross section (and, hence, better limits) for lower 
energies. Thus, a machine operating at $\sqrt{s} = 350 $ GeV would 
do even better for certain couplings than the present case. This is
in agreement with the findings of Ref.~\cite{Biswal:2005fh}. 
For observables constructed with the $R2$-cut, 
however, one can obtain an improvement in the sensitivity limit (say)
for $\Re(\tilde b_Z)$ by upto a factor of 2 at c.m. energy 1 TeV 
compared to the ones possible at 500 GeV \cite{Biswal:2005fh}.

In conclusion, we find that use of beam polarisation and final state
$\tau$ polarisation can result in significant advances 
in the search for anomalous $ZZH$ couplings at 
the ILC. For $WWH$ couplings, though there is
no significant change in the sensitivity limits, the $\tilde T$-even 
ones can be now probed virtually independent of the anomalous $ZZH$ couplings.
The inclusion of ISR and beamstrahlung effects, changes the
individual cross sections but has very little effect 
on the sensitivities. Further, going to higher
beam energies, leads to only modest improvements.
Thus, as far as this sector of the theory is concerned, 
there is a strong case for use of beam polarisation and measurement
of final state fermion polarisations wherever possible, but there is no
real gain in going to higher energies.

\section*{Acknowledgments}
We thank R. K. Singh for collaboration in the initial 
phase of this project. 
DC \& M acknowledge partial support from the Department of
Science and Technology(DST), India under grants 
SR/S2/RFHEP-05/2006 and SR/S2/HEP-12/2006 respectively
and infrastructural support from the IUCAA
Reference Centre, Delhi. 
R.M.G. wishes to acknowledge the DST under Grant No. SR/S2/JCB-64/2007. 
RMG also wishes to acknowledge support from the
Indo French Centre for 
Promotion of  Advanced Scientific Research under project number 3004-2.

\section*{}
{\bf{Note:}} During the last stages of finalising our manuscript, a paper 
dealing with related issues containing a very 
detailed analysis~\cite{sd-hagiwara:2008} on probing the coefficients of the  
CP-even, dimension 6 operators in the anomalous $VVH$ vertices has appeared.

\end{document}